\documentclass[twoside]{article}
\usepackage[latin1]{inputenc}
\usepackage{t1enc}
\usepackage{a4}
\usepackage{tabularx}
\usepackage{epsfig}
\usepackage{xspace}

\def\cref#1{Chapt.\,\ref{#1}}
\def\Cref#1{Chapter~\ref{#1}}
\def\sref#1{Sect.\,\ref{#1}}

\def\fref#1{Fig.\,\ref{#1}}
\def\ffref#1{Figs.\,\ref{#1}}

\def\tref#1{Table~\ref{#1}}
\def\eref#1{(\ref{#1})}

\def\lleft{\textit{left}}
\def\rright{\textit{right}}
\def\LLeft{\textit{Left}}
\def\RRight{\textit{Right}}

\def\TTop{\textit{Top}}
\def\BBottom{\textit{Bottom}}

\def\1{\footnotemark[1]}
\def\and{\& }
\def\Cerenkov{\v{C}erenkov\xspace}

\def\deg{$^\circ$\xspace}

\def\gcm2{g/cm$^2$\xspace}

\def\modell{poly-gonato model\xspace}

\def\Xmax{$X_{max}$\xspace}
\def\PAO{Pierre Auger Observatory\xspace}
\def\uhe{ultra high-energy\xspace}
\def\UHECR{\uhe cosmic rays\xspace}

\def\line{---}
\def\dashed{-\,-\,-}
\def\dotted{$\cdot\cdot\cdot$}
\def\dashdot{$\cdot-\cdot$}

\begin{document}

\setcounter{figure}{0}
\setcounter{section}{0}
\setcounter{equation}{0}

\begin{center}
{\Large\bf
Astronomy with ultra high-energy particles
}\\[0.7cm]

J\"org R. H\"orandel\\[0.17cm]
Department of Astrophysics, Radboud University Nijmegen\\
P.O. Box 9010, 6500 GL Nijmegen, The Netherlands\\
J.Horandel@astro.ru.nl, http://particle.astro.kun.nl\\
\end{center}

\vspace{0.5cm}

\begin{abstract}
\noindent{\it
 Recent measurements of the properties of cosmic rays above $10^{17}$~eV are
 summarized and implications on our contemporary understanding of their origin
 are discussed.  Cosmic rays with energies exceeding $10^{20}$~eV have been
 measured, they are the highest-energy particles in the Universe.  Particles at
 highest energies are expected to be only marginally deflected by magnetic
 fields and they should point towards their sources on the sky.  Recent results
 of the \PAO have opened a new window to the Universe --- astronomy with \uhe
 particles.
}
\end{abstract}

\section{Introduction}

The Earth is permanently exposed to a vast flux of high-energy particles from
outer space. Most of these particles are fully ionized atomic nuclei with
relativistic energies.  The extraterrestrial origin of these particles has been
demonstrated by V. Hess in 1912 \cite{hess} and he named the particles
"H\"ohenstrahlung" (high-altitude radiation) or "Ultrastrahlung" (ultra
radiation). In 1925 R. Millikan coined the term "Cosmic Rays". They have a
threefold origin.  Particles with energies below 100~MeV \footnote{In this
review we use the particle physics energy units MeV$=10^{6}$~eV,
GeV$=10^{9}$~eV, TeV$=10^{12}$~eV, PeV$=10^{15}$~eV, and EeV$=10^{18}$~eV;
1~eV$=1.6\cdot10^{-19}$~J.  } originate from the Sun
\cite{ryanpune,kahlerpune}.  Cosmic rays in narrower sense are particles with
energies from the 100~MeV domain up to energies beyond $10^{20}$~eV. Up to
several 10~GeV the flux of the particles observed is modulated on different
time scales by the heliospheric magnetic fields \cite{fichtnerpune,heberecrs}.
Particles with energies below $10^{17}$ to $10^{18}$~eV are usually considered
to be of galactic origin
\cite{gaisserstanev,gaisserjapan,smparnp,cospar06,ricap07}. The Larmor radius
of a particle with energy $E_{15}$ (in units of $10^{15}$~eV) and charge $Z$ in
a magnetic field $B_{\mu {\rm G}}$ (in $\mu$G) is
\begin{equation}\label{larmor}
 r_L=1.08 \, \frac{E_{15}}{Z B_{\mu {\rm G}}}~\mbox{pc} ,
\end{equation}
yielding a value of $r_L=360$~pc for a proton with an energy of $10^{18}$~eV in
the galactic magnetic field ($B_{\mu{\rm G}}\approx3$).  This radius is
comparable to the thickness of the galactic disc and illustrates that particles
(at least with small charge $Z$) at the highest energies can not be
magnetically bound to the Galaxy.  Hence, they are considered to be of
extragalactic origin \cite{naganowatson,bergmanbelz,kamperttaup07}. 

In the present article, we focus on recent results concerning the origin of the
extragalactic particles. Cosmic rays with energies exceeding $10^{20}$~eV are
the highest-energy particles in the Universe.  Particles at highest energies
are only marginally deflected in the galactic magnetic fields, following
\eref{larmor} they have a Larmor radius $r_L>36$~kpc, exceeding the diameter of
the Milky Way.  Thus, they should point back to their sources, enabling
astronomical observations with charged particles.  

Several questions arise, concerning the origin of highest-energy cosmic rays.
Among them are:\\
-- What are the energies of the particles? (\sref{energy})\\
-- What are these particles? Are they protons, nuclei of heavy atoms like
oxygen or iron, furthermore are they photons or neutrinos? (\sref{mass})\\
-- Where do they come from? Can we learn something by studying their arrival
directions? (\sref{direction})\\
-- How do they propagate to us? Do they suffer any interactions?
(\sref{energy})

In the following sections (\ref{energy} to \ref{direction}) recent
experimental results are compiled and their implications to answer the
questions raised above are discussed. 
Before, possible scenarios for the sources of the particles are summarized
(\sref{sources}) and mechanisms are discussed which are important during the
propagation of the particles through the Universe (\sref{propagation}).
The detections methods applied are sketched in \sref{detection}.

\section{Sources and Propagation}\label{sup}

\subsection{Sources}\label{sources}

The energy density contained in the flux of extragalactic cosmic rays
can be inferred from the measured differential energy spectrum $dN/dE$
\cite{halzenrhoe}
\begin{equation}
 \rho_E=\frac{4\pi}{c}\int \frac{E}{\beta} \frac{dN}{dE} dE ,
\end{equation}
where $\beta c$ is the velocity of particles with energy $E$.  
To estimate the energy content of the extragalactic component, assumptions have
to be made about the contribution of galactic cosmic rays at energies in the
transition region ($10^{17}-10^{18}$~eV). The extragalactic component needed
according to the \modell \cite{pg} to sustain the observed all-particle flux at
highest energies has an energy density of $\rho_E=3.7\cdot10^{-7}$~eV/cm$^3$.
The power required for a population of sources to generate this energy density
over the Hubble time of $10^{10}$ years is $5.5\cdot10^{37}$~erg/(s Mpc$^3$).
This leads to $\approx2\cdot10^{44}$~erg/s per active galaxy or
$\approx2\cdot10^{52}$ erg per cosmological gamma ray burst
\cite{gaisser-eden}. The coincidence between these numbers and the observed
output in electromagnetic energy of these sources explains why they are
considered as promising candidates to accelerate highest-energy cosmic rays.

\begin{figure}
 \epsfig{width=0.6\textwidth, file=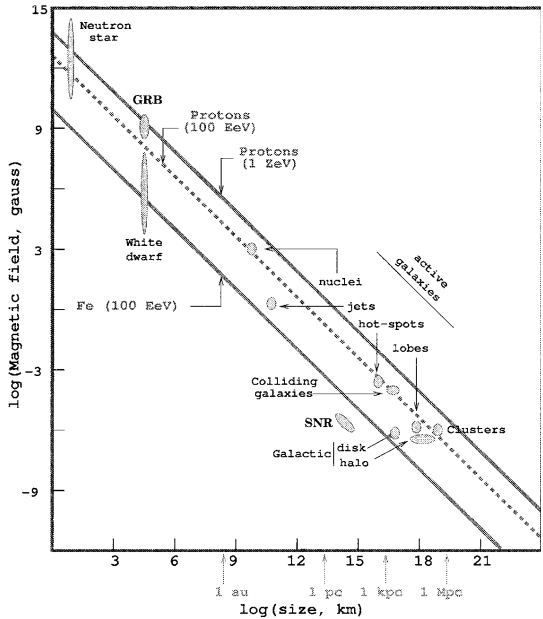}\hspace{\fill}
 \begin{minipage}[b]{0.37\textwidth}
 \caption{Size and magnetic field strength of possible sites of particle
	  acceleration (Hillas diagram). Acceleration of cosmic rays up to a
	  given energy requires conditions above the respective line
	  \cite{ostrowskihillas}.\label{hillas}}
 \end{minipage}
\end{figure}

The characteristic size of an accelerating region can be estimated for models
of gradual acceleration, where the particles make many irregular loops in a
magnetic field while gaining energy \cite{hillasdiagram}. The size $L$ of the
essential part of the accelerating region containing the magnetic field must be
grater than $2r_L$. A closer look reveals that a characteristic velocity $\beta
c$ of scattering centers is of virtual importance \cite{hillasdiagram}, which
yields the expression
\begin{equation}\label{hillascond}
 B_{\mu{\rm G}} L_{\rm pc} > 2 E_{15}/(Z\beta) .
\end{equation}
It relates the characteristic size $L_{\rm pc}$ (in pc) and magnetic fields
$B_{\mu{\rm G}}$ of objects being able to accelerate particles to energies
$E_{15}$.  Several possible acceleration sites are explored in \fref{hillas},
where the magnetic field strength is plotted as function of their typical sizes
\cite{ostrowskihillas}. The lines according to \eref{hillascond} represent
the conditions for protons and iron nuclei of different energies, as indicated.
Objects capable to accelerate particles above a respective energy should be
above the respective line. As can be inferred from the figure, most promising
candidates to accelerate highest-energy cosmic rays are gamma ray bursts and
active galactic nuclei (AGN) \cite{ginzburg,hillasdiagram}. 
These objects are typically in a distance of several tens of Mpc to the Earth. 
Interactions in the source itself or in the vicinity of the source of hadronic
particles (protons, nuclei) yield neutral and charged pions, which subsequently
decay into high-energy photons and neutrinos.

Alternatively, so called "top-down models" are discussed in the literature
\cite{hillschramm,olintopr,siglpr}.  They have been motivated by events seen by
the AGASA experiment above the threshold for the GZK effect \cite{agasagzk}.
It is proposed that \uhe particles (instead of being accelerated,
"bottom-up scenario") are the decay products of exotic, massive particles
originating from high-energy processes in the early Universe.  Such
super-massive particles (with $m_X \gg 10^{11}$~GeV) decay e.g.\ via W and Z
bosons into high-energy protons, photons, and neutrinos.

\subsection{Propagation}\label{propagation}
On the way from their sources to Earth the particles propagate mostly outside
galaxies in intergalactic space with very low particle densities.  In this
environment the most important interactions of cosmic rays occur with photons
of the 2.7-K microwave background radiation, namely pair production and pion
photoproduction \cite{hill-schramm}. 

On the last part of their way to Earth they propagate through the Galaxy.
However, since particles at the highest energies travel almost along straight
lines they accumulate a negligible amount of material during their short
travel through regions with relatively high densities. Thus, interactions with
the interstellar material can be neglected.

The Universe is filled with about 412~photons/cm$^3$ of the 2.7\deg K microwave
background radiation.  Shortly after the discovery of the microwave background
it was proposed that \UHECR should interact
with the photons, leading to a suppression of the observed flux at highest
energies \cite{gzkgreisen,gzkzatsepin}. This effect is called after its
proposers the Greisen-Zatsepin-Kuz'min (GZK) effect. A nucleon of energies
exceeding $E_{GZK}\approx6\cdot10^{19}$~eV colliding head-on with a 2.7\deg K
photon comprises a system of sufficient energy to produce pions by the
photoproduction reaction 
\begin{equation}\label{gzkeq}
  p + \gamma_{3K}\rightarrow \Delta^+ \rightarrow 
  \begin{array}{l} p+\pi^0 \\ n+\pi^+ \end{array} .
\end{equation}
The energy loss of the nucleon is a significant fraction of the initial energy.
The pion photoproduction cross section is quite large above threshold due to
resonance production ($\Delta$ resonance), rising quickly to 500~mb for photon
laboratory energies of about 0.3~GeV.  Subsequent decays of the neutral and
charged pions produced in \eref{gzkeq} yield high-energy photons and neutrinos.

The center of mass energy of interactions of cosmic rays with energies
exceeding $10^{18}$~eV colliding with microwave-background photons is 
sufficient to generate electron-positron pairs $p+\gamma_{3K}\rightarrow
p+e^++e^-$. As a consequence, the cosmic-ray particles loose energy which
leads presumably to a reduction of the flux or a dip in the spectrum between
$10^{18}$ and $10^{19}$~eV \cite{hill-schramm,berezinskydip,berezinskydipapp}.

\begin{figure}\centering
  \epsfig{width=0.53\textwidth, file=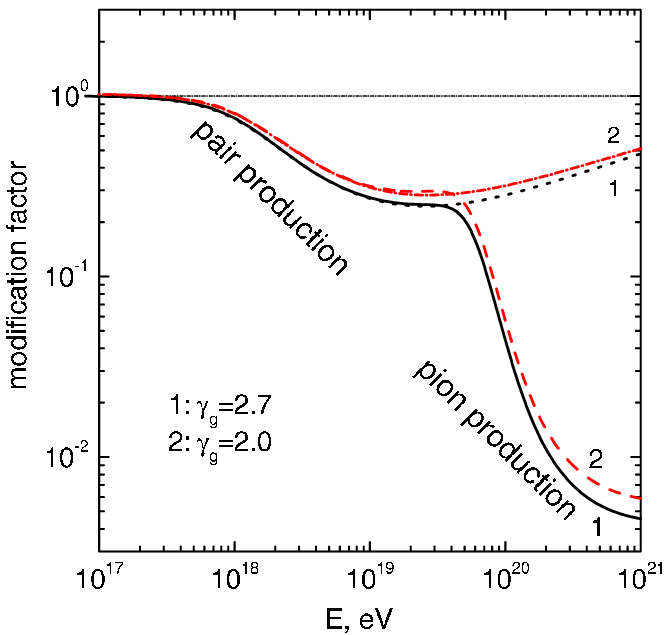}\hspace*{\fill}
  \epsfig{width=0.45\textwidth, file=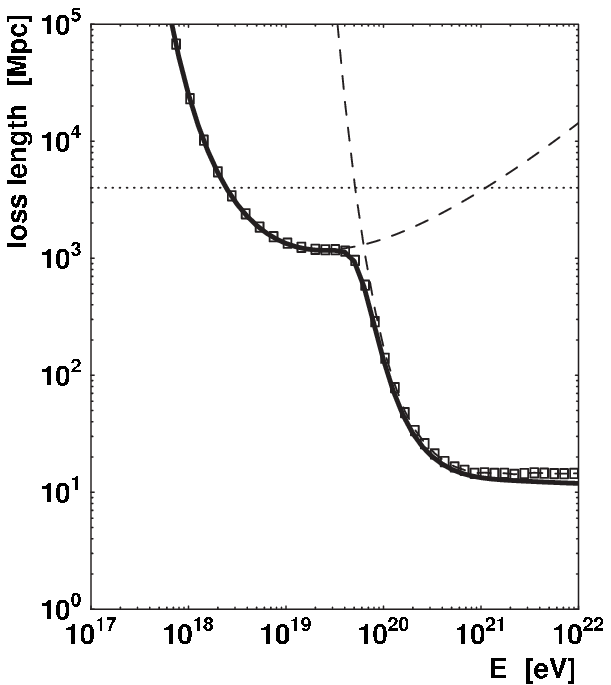}
  \caption{\LLeft: Modification factor $f(E)=I_p(E)/I_0(E)$ of the cosmic-ray
	   energy spectrum \cite{dipapp}.
           \RRight: Loss length for protons for pair production and pion
           photoproduction \cite{demarco-stanev}.}
 \label{dipgzk}	  
\end{figure} 

The effect of both processes on the observed energy spectrum is frequently
expressed by a modification factor $f(E)=I_p(E)/I_0(E)$, describing the ratio
of the observed spectrum $I_p$ and the initial spectrum $I_0$ as function of
energy. The modification factor according to recent calculations is shown in
\fref{dipgzk} (\lleft) \cite{dipapp}. A twofold structure can be recognized. A
depression (the dip) at energies between $10^{18}$ and $10^{19}$~eV and the GZK
feature at energies exceeding $5\cdot10^{19}$~eV. The two cases (1 and 2)
represent initial spectra with a spectral index of 2.0 and 2.7, respectively.

The energy loss length for pair production and pion photoproduction is
depicted in \fref{dipgzk} (\rright) \cite{demarco-stanev}.  Particles with
energies above the threshold of the GZK effect can travel less than about
100~Mpc through the Universe, before their energy has decreased to $1/e$ of
their initial value. Or, in other words, particles reaching the Earth at these
energies have propagated less than 100~Mpc, see also \cite{cronin-gzk},
and their sources are inside a sphere with this radius

\subsection{Multi Messenger Approach}
It has been discussed that for both scenarios, acceleration and top-down
models, hadronic cosmic rays are accompanied by high-energy photons and
neutrinos. Also during the propagation of hadronic particles through the
Universe high-energy photons and neutrinos are produced. To clarify the origin
of the highest-energy particles in the Universe, simultaneous observations are
desired of high-energy charged particles, photons, and neutrinos --- a
multi-messenger approach.  Thus, the observation of high-energy charged
particles, or charged particle astronomy, is complementary to observations in
gamma ray astronomy \cite{ong,ongpune} and neutrino astronomy
\cite{spieringecrs,halzenheraeus,berezinskypylos,liparicatania}.

Attention has to be paid on the 'simultaneous' observation: if a charged
particle is deflected by an angle $\Theta$ in a (for this estimate simply
homogeneous) magnetic field, its path $L_{ch}$ is somewhat longer as compared
to the path of a massless neutral particle $L_\gamma$. Their relative
difference can be approximated as
\begin{equation} {\cal R}=
 \frac{L_{ch}}{L_\gamma} = \frac{2 \pi \Theta}{360^\circ \sin(\Theta)} .
\end{equation}
If a charged particle from a source at a distance $L_\gamma=100$~Mpc is
deflected by $\Theta=3^\circ$, a value ${\cal R}=1+4.6\cdot10^{-4}$ is
obtained. This corresponds to a difference in the arrival time of a charged
particle relative to a photon (both traveling at the speed of light) of about
$\Delta T=150\cdot10^3$~a. Thus, for a simultaneous detection the acceleration
processes have to be stable over such a period in time.

\section{Detection Method} \label{detection}
The extremely steeply falling cosmic-ray energy spectrum ($\propto E^{-3}$)
yields very low fluxes for the highest-energy particles. At the highest
energies less than one particle is expected per square kilometer and century.
This necessitates huge detection areas and large measuring times. At present,
they are only realized in huge ground based installations, registering
secondary particles produced in the atmosphere.

\subsection{Extensive air showers} 

When high-energy cosmic-ray particles penetrate the Earths atmosphere they
interact and generate a cascade of secondary particles, the extensive air
showers. Hadronic particles interact and produce new hadronic particles or
generate muons and photons through pion decays. Some of the muons may decay
into electrons, while the photons and electrons/positrons regenerate themselves
in an electromagnetic cascade.  The by far dominant particles in a shower are
electromagnetic particles (photons, electrons, and positrons).  Most of the
energy of the primary particle is absorbed in the atmosphere. However, a small
fraction of the energy is transported to ground level and may be registered in
detectors for electrons, muons, and hadrons.  Particles traveling with
relativistic speeds through the atmosphere (mostly electrons and positrons)
emit \Cerenkov light. The shower particles also excite nitrogen molecules in
the air which in turn emit fluorescence light.  While the \Cerenkov light is
collimated in the forward direction of the particle, the fluorescence light is
emitted isotropically, thus, a shower can be "viewed from aside". 

The objective of experiments observing extensive air showers is to determine
the properties of the primary particle (energy $E_0$, mass $A$, arrival
direction).  In the energy regime of interest ($E>10^{17}$~eV) mainly two
methods are applied. Electrons (and positrons) as well as muons reaching ground
level are observed in large arrays of detectors and the fluorescence light is
viewed by imaging telescopes.
An alternative technique, presently under investigation, is the detection of
radio emission from air showers.  Electrons and positrons are deflected in the
Earths magnetic field and emit synchrotron radiation, which is detected in
arrays of dipole antennae \cite{radionature,vdbergmerida,huegefalcke}.

\subsection{Measuring technique}
The {\bf direction} of air showers is inferred applying two techniques.  The
particles in a shower travel with nearly the speed of light through the
atmosphere in a thin disc with a thickness of a few meters only.  With
detectors measuring the arrival time of the particles with a resolution of a
few ns the angle of the shower front relative to the ground can be inferred,
with the arrival direction being perpendicular to the shower plane.  With
imaging fluorescence telescopes the shower-detector plane is determined from
the observed track in the camera. The orientation of the shower axis in this
plane is then obtained by measurements of the arrival time of the photons at
the detector.  Using two (or more) telescopes to view the same shower allows a
three-dimensional reconstruction of the shower axis.

The shower {\bf energy} is proportional to the number of electrons $N_e$ and
muons $N_\mu$ in the shower. A simple numerical model \cite{jrherice06} yields
the relations
\begin{equation}
 E_0=3.01~\mbox{GeV}\cdot A^{0.04} \cdot N_e^{0.96}
 \quad\mbox{and}\quad
 E_0=20~\mbox{GeV}\cdot A^{-0.11} \cdot N_\mu^{1.11}
\end{equation}
to estimate the primary energy.  This illustrates that measuring $N_e$ or
$N_\mu$ gives a good estimate for the energy almost independent of the
particles mass.\\
With imaging fluorescence telescopes the amount of fluorescence light can be
measured as function of depth in the atmosphere.  The total amount of light
collected is proportional to the shower energy.
Using the number of electrons at shower maximum, the number of photons
registered per square meter in a detector at a distance $r$ to the maximum of a
shower with energy $E_0$ can be estimated as
\begin{equation}\label{photoneq}
 \nu_\gamma=\frac{N_e X_0 N_\gamma}{4\pi r^2}
  \approx 790\,\frac{\gamma}{\mbox{m}^2} \, A^{-0.046} \,
                \left(\frac{E_0}{\mbox{EeV}}\right)^{1.046} 
		\frac{1}{\left(r/10~\mbox{km}\right)^2} ,
\end{equation}
where $N_\gamma\approx 4~\gamma/\mbox{m}$ is the fluorescence yield of
electrons in air and $X_0=36.7$~\gcm2 (or 304~m at normal pressure) the
radiation length.  Absorption and scattering in the atmosphere have been
neglected in this simple estimate, thus, the equation gives an upper limit for
the registered photons.

Experimentally most challenging is the estimation of the {\bf mass} of the
primary particle. Showers induced by light and heavy particles develop
differently in the atmosphere. The depth in the atmosphere \Xmax at which the
showers contain a maximum number of particles depends on the primary particles
mass
\begin{equation} \label{xmaxeq}
 X_{max}^A=X_{max}^p-X_0\ln A,
\end{equation}
where $X_{max}^p$ is the depth of the shower maximum for proton-induced showers
\cite{matthewsheitler,jrherice06}. Experiments measuring the longitudinal
shower profile by observations of fluorescence light estimate the mass by
measurements of \Xmax.\\
If a shower develops higher in the atmosphere more particles (mostly electrons)
are absorbed on the way to the ground. On the other hand, at high altitudes
(with low air densities) charged pions are more likely to decay, thus, yielding
more muons.  Hence, the electron-to-muon ratio observed at ground level depends
on the mass of the primary particle.  A Heitler model of hadronic showers
\cite{jrherice06} yields the relation
\begin{eqnarray} \label{emratioeq}
 \frac{N_e}{N_\mu}\approx35.1
     \left(\frac{E_0}{A\cdot1~\mbox{PeV}}\right)^{0.15} .
\end{eqnarray}
This implies that the registered electron-to-muon ratio depends on the energy
per nucleon of the primary particle.

\subsection{Cosmic-Ray Detectors}

In the following we describe the most important recent detectors for \UHECR.

\paragraph{The AGASA experiment}
The Akeno Giant Air Shower Array (AGASA) was a scintillator array located in
Japan (35$^\circ$N, 138$^\circ$E), covering an area of 100~km$^2$ \cite{agasa}.
It consisted of 111 scintillation counters to register the electromagnetic
shower component.  Each station covered 2.2~m$^2$ in area.  The scintillator
blocks with a thickness of 5~cm were viewed by a 125~mm diameter
photomultiplier tube.  To register the muonic shower component, proportional
counters were used with a cross section of $10\times10$~cm$^2$ and a length of
2~m or 5~m. The absorber consisted either of a 1~m thick concrete block, a
30~cm thick iron plate, or a 5~cm lead plate above a 20~cm thick iron plate.
The threshold energy for muons is about 0.5~GeV. In total, 27 detector stations
were installed with areas varying from 2.8~m$^2$ to 20~m$^2$.

\paragraph{The HiRes experiment}
The High Resolution Fly's Eye experiment (HiRes) was located in Utah, USA
(40$^\circ$N, 112$^\circ$W) \cite{hiresexp}. It was the successor of the Fly's
Eye experiment \cite{flyseyeexp}, which pioneered the detection of fluorescence
light from air showers.  HiRes consisted of two detector sites (Hires I \& II)
separated by 12.6~km, providing almost 360\deg azimuthal coverage, each.  Both
telescopes were formed by an array of detector units. The mirrors consisted of
four segments and formed a 5.1~m$^2$ spherical mirror. At its focal plane an
array of $16\times16$ photomultiplier tubes was situated, viewing a solid angle
of $16^\circ\times16^\circ$.  HiRes~I consisted of 22 detectors, arranged in a
single ring, overlooking between 3\deg and 17\deg in elevation. This detector
used an integrating ADC read-out system, which recorded the photomultiplier
tubes' pulse height and time information.  HiRes~II comprises 42 detectors, set
up in two rings, looking between 3\deg and 31\deg in elevation. It was equipped
with a 10~MHz flash ADC system, recording pulse height and timing information
from its phototubes.

\paragraph{The \PAO}

\begin{figure}[t] \centering
 \epsfig{width=0.49\textwidth, file=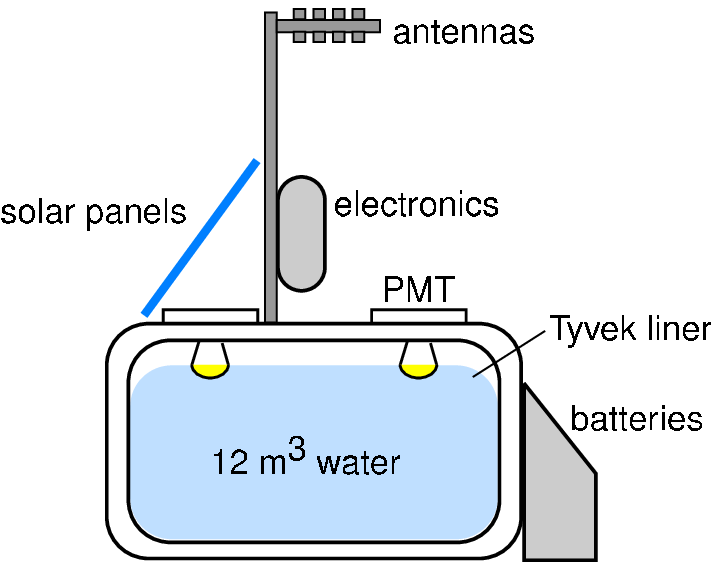}
 \epsfig{width=0.49\textwidth, file=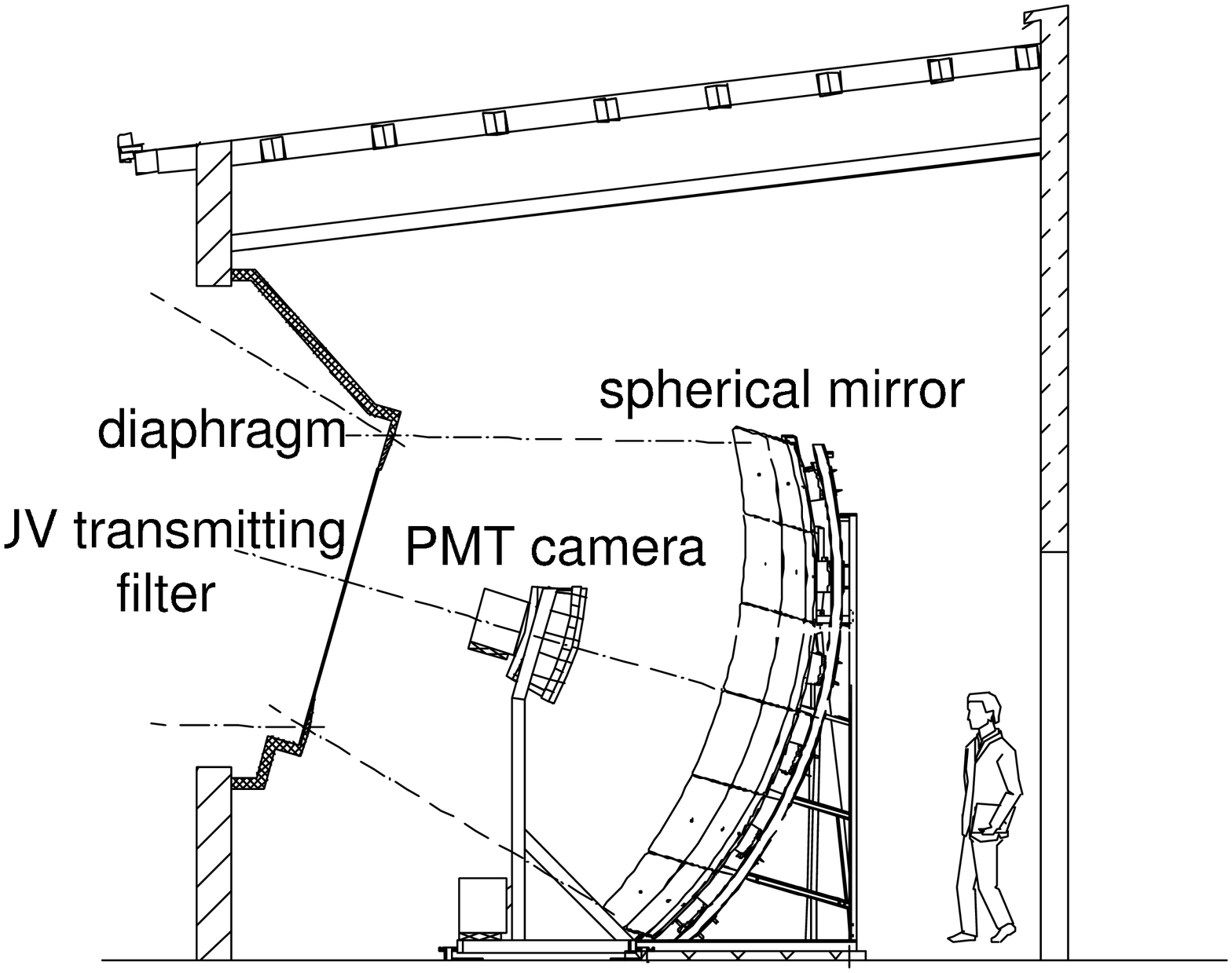}
 \caption{Schematic view of a water \Cerenkov detector (\lleft) and a
	  fluorescence telescope (\rright) of the \PAO \cite{augerexp}.}
 \label{auger}
\end{figure}

The observatory combines the observation of fluorescence light with imaging
telescopes and the measurement of particles reaching ground level in a "hybrid
approach" \cite{augerexp}.  The southern site (near Malargue, Argentina,
35.2\deg S, 69.5\deg W, 1400~m above sea level) of the worlds largest air
shower detector is almost completed.  It will consist of 1600 polyethylene tanks
set up in an area covering 3000~km$^2$. Each water \Cerenkov detector has 3.6~m
diameter and is 1.55~m high, enclosing a Tyvak liner filled with 12~m$^3$ of
high purity water, see \fref{auger}. The water is viewed by three PMTs 
(9~in diameter).  Signals from the PMTs are read by the electronics mounted
locally at each station. Power is provided by batteries, connected to solar
panels, and time synchronization relies on a GPS receiver. A radio system is
used to provide communication between each station and a central data
acquisition system.\\ 
Four telescope systems overlook the surface array.  A single telescope system
comprises six telescopes, overlooking separate volumes of air.  A schematic
cross-sectional view of one telescope is shown in \fref{auger}.  Light enters
the bay through an UV transmitting filter.  A circular diaphragm (2.2~m
diameter), positioned at the center of curvature of a spherical mirror, defines
the aperture of the Schmidt optical system.  A $3.5~\mbox{m}\times3.5~\mbox{m}$
spherical mirror focuses the light onto a camera with an array of $22\times20$
hexagonal pixels.  Each pixel has a photomultiplier tube, complemented by light
collectors.  Each camera pixel has a field of view of approximately 1.5\deg, a
camera overlooks a total field of view of 30\deg azimuth $\times$ 28.6\deg
elevation.

\paragraph{Telescope Array}
Like the \PAO, the Telescope array is a hybrid detector, presently under
construction in Mullard County, Utah, USA \cite{telescopearray}. It covers an
area of 860~km$^2$ and comprises 576 scintillator stations and three
fluorescence detector sites on a triangle with about 35~km separation, each
equipped with twelve fluorescence telescopes.

\paragraph{}

\begin{figure}
 \epsfig{width=0.6\textwidth, file=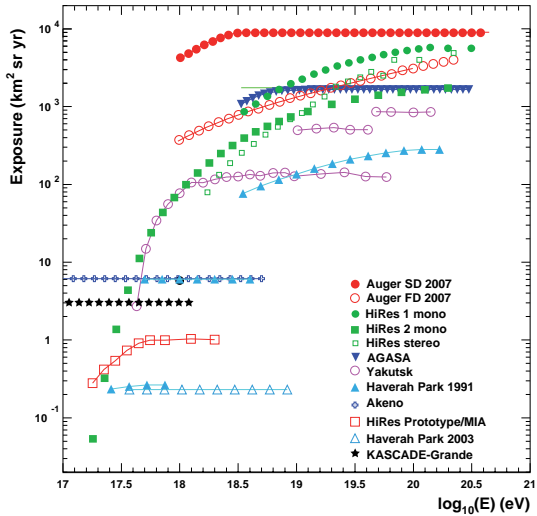}\hspace{\fill}
 \begin{minipage}[b]{0.37\textwidth}
 \caption{Accumulated exposures of various experiments
	  \cite{kamperttaup07}.\label{apperture}}
 \end{minipage}
\end{figure}

Accumulated exposures (i.e.\ experiment aperture times measuring time) for
various high-energy experiments are presented in \fref{apperture}
\cite{kamperttaup07}.  For surface arrays the aperture is a function of the
detector area and constant with energy. On the other hand, the aperture of
fluorescence detectors depends on the shower energy, low energy showers can be
seen up to a restricted distance only. This may be illustrated using the
approximation \eref{photoneq}: the fiducial volume to register
$\nu_\gamma^{min}$ photons can be estimated as 
\begin{equation}
 V_{fid} \propto (\nu_\gamma^{min})^{-1.5} \, A^{-0.069} \, E_0^{1.569}. 
\end{equation}
This shows that the fiducial volume is a function of the primary energy.  In
this simple approximation there is a small dependence on the mass of the
primary particle ($\approx25\%$ difference between proton and iron induced
showers) and an increase of almost a factor of 40 in the fiducial volume per
decade in primary energy. A similar energy dependence can be recognized in
\fref{apperture} for the various fluorescence detectors. For fluorescence
telescopes with a limited field of view in elevation an additional effect
occurs: low energy showers penetrate less deep into the atmosphere and thus
may have their maximum above the field of view of the telescopes, thus,
reducing further the effective aperture. 

The \PAO, still under construction, is already the largest cosmic ray detector,
the accumulated data exceed the data of all previous experiments.  In
particular, those of the largest scintillator array (AGASA) and the largest
pure fluorescence detector (HiRes).  Thus, the \PAO is expected to measure the
properties of \UHECR with unprecedented accuracy.

\section{Energy Spectrum}\label{energy}

Measurements of the energy spectrum provide important information about the
origin of cosmic rays.  Over a wide range in energy the all-particle
differential energy spectrum is usually described by a power law $dN/dE\propto
E^{-\gamma}$.  For energies below $10^{15}$~eV a value for the spectral index
$\gamma=-2.7$ has been established by many experiments.  The most prominent
feature in the all-particle spectrum is the so called knee at an energy of
about $4\cdot10^{15}$~eV. At this energy the spectral index changes to
$\gamma\approx-3.1$.  The knee in the all-particle spectrum is caused by the
subsequent cut-offs (or knees) of the spectra of individual elements, starting
with protons at $E_k^p\approx4.5\cdot10^{15}$~eV.  However, this feature is
below the focus of the current article, thus, the reader may be referred to
e.g.\ \cite{pg,origin,cospar06} for a more detailed discussion about galactic
cosmic rays and the knee.  In the following we focus on energies above
$10^{17}$~eV.

\begin{figure}
 \epsfig{width=0.6\textwidth, file=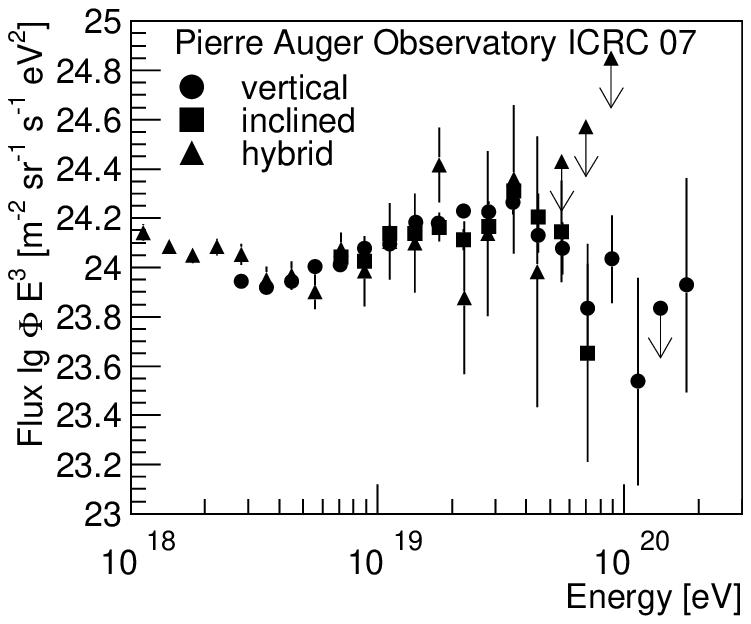}\hspace{\fill}
 \begin{minipage}[b]{0.37\textwidth}
 \caption{All-particle energy spectra measured by the \PAO 
	  using different reconstruction methods \cite{auger07}.\label{augere}}
 \end{minipage}
\end{figure}

Recent energy spectra as obtained by the \PAO are depicted in \fref{augere}
\cite{auger07}. The registered flux has been multiplied by $E^3$. Different
methods are applied to reconstruct the spectra.  The first method uses the data
from the 3000~km$^2$ surface array. The detection efficiency reaches 100\% for
showers with zenith angles less than 60\deg for energies above $10^{18.5}$~eV
and for inclined showers ($\Theta>60^\circ$) above $10^{18.8}$~eV. The signal
at 1000~m from the shower axis is used to estimate the shower energy. To avoid
a dependence on interaction models used in air shower simulation codes, an
energy estimator is derived based on measured showers: a subset of showers
contains so called hybrid events, seen simultaneously by the surface detector
array and at least one fluorescence telescope. The fluorescence telescopes
provide a nearly model independent calorimetric energy measurement of the
showers in the atmosphere. Only a small correction for 'invisible energy'
(high-energy muons and neutrinos carrying away energy) has to be applied. This
factor amounts to about 10\% and contributes with about 4\% to the systematic
error for the energy.  The energy calibration thus obtained is applied to all
events recorded with the surface detector array.  Also inclined events with
zenith angles exceeding 60\deg have been analyzed, yielding the second spectrum
displayed.  Finally, a set of showers which have been recorded by at least one
surface detector tank and one fluorescence telescope have been analyzed. The
resulting energy spectrum reaches energies as low as $10^{18}$~eV, as can be
inferred from \fref{augere}.  It is interesting to point out that the different
spectra have been analyzed independently and agree quite good with each other.

\begin{table}\centering
 \caption{Energy shifts applied to individual experiments as shown in
          \fref{espek}.\label{etab}}
 \begin{tabular}{llr} \hline
  Experiment & Reference & Energy shift \\ \hline
  AGASA                 & \cite{agasa}          &  -22\%   \\
  Akeno 1 km$^2$        & \cite{akeno1}         &  -4\%    \\
  Akeno 20 km$^2$       & \cite{akeno20}        &  -22\%   \\
  Auger                 & \cite{auger07}        &  +20\%   \\
  Fly's Eye             & \cite{flyseye}        &  -3\%    \\
  Haverah Park          & \cite{haverahpark03}  &  -2\%    \\
  HiRes-I               & \cite{hiresi}         &  0\%     \\
  HiRes-II              & \cite{hiresii}        &  0\%     \\
  HiRes-MIA             & \cite{hiresxmax}      &  +5\%    \\
  KASCADE-Grande        & \cite{kgvanburen}     &  -7\%    \\
  MSU                   & \cite{msu}            &  -5\%    \\
  SUGAR                 & \cite{sugar}          &   0\%    \\
  Yakutsk T500          & \cite{yakutsk5001000} & -35\%    \\
  Yakutsk T1000         & \cite{yakutsk5001000} & -20\%    \\
  \hline
 \end{tabular}
\end{table}

\begin{figure}\centering
 \epsfig{width=0.95\textwidth, file=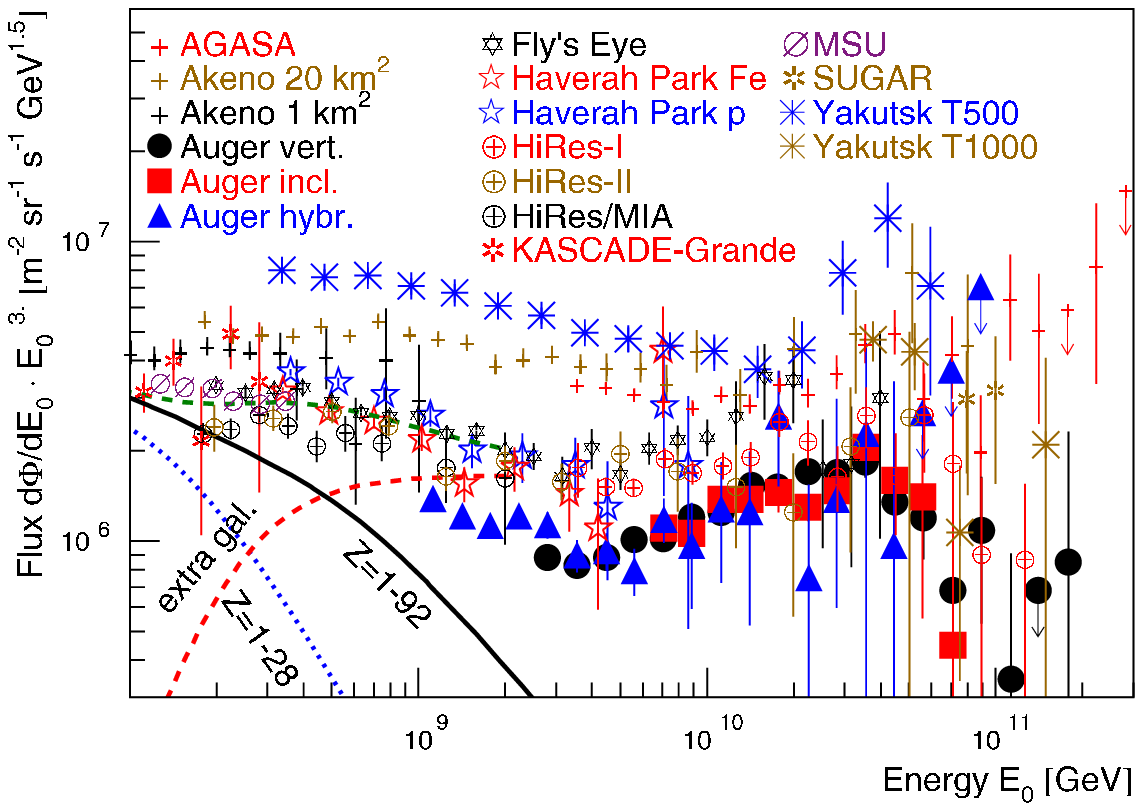}
 \epsfig{width=0.95\textwidth, file=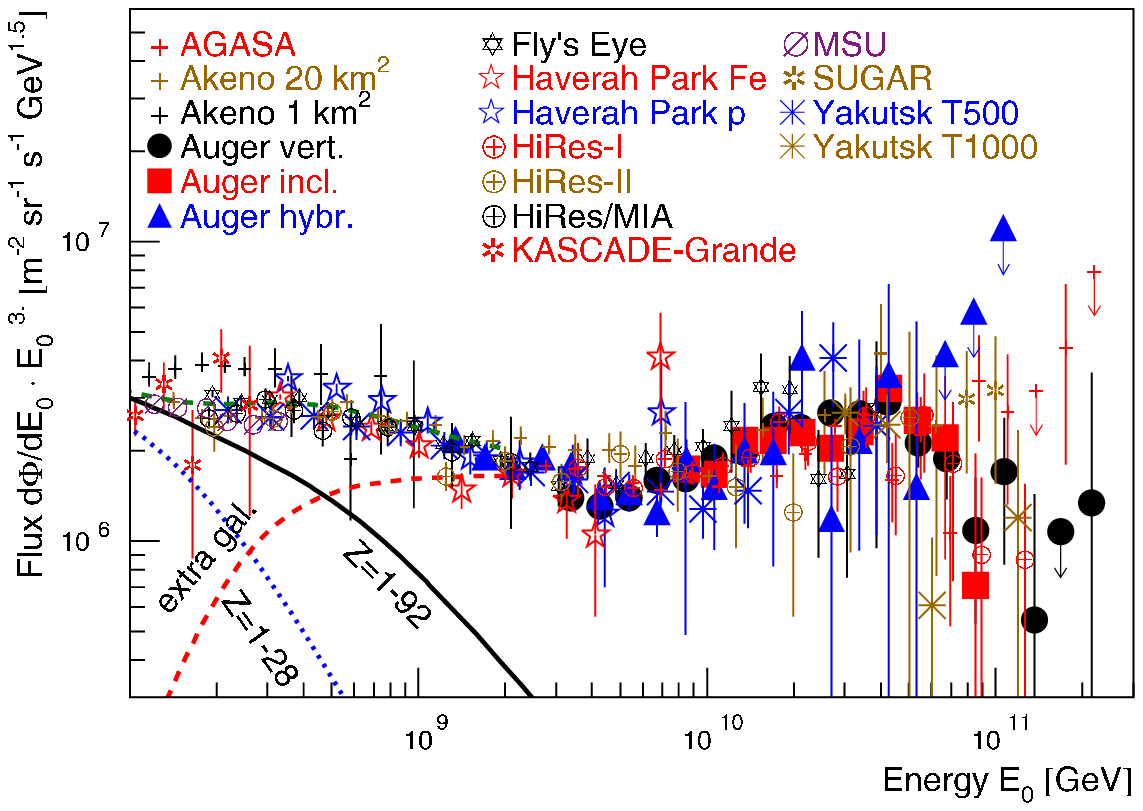}
 \caption{All-particle energy spectra as obtained by different experiments.
	  The top panel shows the original values, in the bottom panel the
	  energy scales of the individual experiments have been adjusted. For
	  references and energy shifts, see \tref{etab}. The lines indicate the
	  end of the galactic component according to the \modell \cite{pg} and
	  a possible contribution of extragalactic cosmic rays.
	  \label{espek}}
\end{figure}

The all-particle energy spectra as obtained by various experiments are compiled
in \fref{espek}. The flux has been multiplied by $E^3$. The upper panel shows
the original data.  The different experiments yield absolute values which
differ by almost an order of magnitude in this representation. Nevertheless,
the overall shape of the energy spectrum seems to be reflected in all data,
irrespective of the absolute normalization. This becomes more obvious when the
energy scales are slightly readjusted. Typical systematic uncertainties for the
energy scale are of order of 10\% to 30\% in the region of interest.  When
energy shifts are applied, the results have to be treated with care since the
apertures of some experiments change as function of energy (see
\fref{apperture}) and this effect has not been taken into account in the
procedure used here.

In the lower panel of \fref{espek} the energy scales of the different
experiments have been adjusted to fit the flux according to the \modell at
$10^{18}$~eV. The latter has been obtained through a careful procedure
extrapolating the measured spectra for individual elements at low energies
\cite{pg}. Thus, the normalization applied provides a consistent description
from direct measurements (10~GeV region) up to the highest energies. The
corresponding energy shifts are listed in \tref{etab}.

The normalized spectra agree very well and seem to exhibit a clear shape of the
all-particle energy spectrum. Some structures seem to be
present in the spectrum. The second knee at about $4\cdot10^{17}~$eV, where the
spectrum steepens to $\gamma\approx-3.3$ and the ankle at about
$4\cdot10^{18}$eV, above this energy the spectrum flattens again to
$\gamma\approx-2.7$. Finally, above $4\cdot10^{19}$~eV the spectrum exhibits
again a steepening with a spectral index $\gamma\approx-4$ to $-5$.  The new
Auger results help to clarify the situation in this energy region.  While the
AGASA experiment has reported events beyond the GZK threshold \cite{agasagzk},
the HiRes experiment has reported a detection of the GZK cut-off
\cite{hiresgzk}. With the new results, a steeper falling spectrum above
$4\cdot10^{19}$~eV is now confirmed.

The second knee possibly marks the end of the galactic component \cite{pg}.  If
the energy spectra for individual elements exhibit knees at energies
proportional to their nuclear charge, the heaviest elements in galactic cosmic
rays should fall off at an energy of about $92\cdot
E_k^p\approx4\cdot10^{17}$~eV. An interesting coincidence with the position of
the second knee.  Different scenarios for the transition from galactic to
extragalactic cosmic rays are discussed e.g.\ in \cite{kampertcris06,cospar06}.
In the energy region around the ankle a depression is seen in the all-particle
flux, also referred to as the dip. It is proposed that this dip is caused by
interactions of \uhe particles with the cosmic microwave
background, resulting in electron-positron pair production, see
\sref{propagation}. 
The steepening in the flux above $4\cdot10^{19}$~eV could be an indication of
the GZK effect, i.e. photo-pion production of \UHECR
with the microwave background, see \sref{propagation}.  However, for a definite
answer also other properties of cosmic rays have to be investigated.

\section{Mass Composition}\label{mass}

The elemental composition of galactic cosmic rays has been discussed elsewhere,
e.g.\ \cite{wq,cospar06}. Above $10^{17}$~eV the situation is experimentally
very challenging, since we are far away in parameter space from collider
experiments, where the properties of high-energy interactions are studied in
detail. Thus, the air shower models used to interpret the data have to
extrapolate over a wide range in parameter space.

The fraction of iron nuclei in cosmic rays as deduced by many experiments has
been investigated \cite{dovafefract}.  No clear conclusion can be drawn about
the composition at highest energies.
Tension in the interpretation of the measured data  has been observed as well
by the HiRes-MIA experiment \cite{hires00}. The observed \Xmax values exhibit a
trend towards a lighter composition as function of energy in the range between
$10^{17}$ and $10^{18}$~eV. On the other hand, measured muon densities indicate
a very heavy composition in the same energy range.

Methods relying on the measured muon densities, the lateral distribution of
\Cerenkov light registered at ground level, or geometrically-based methods are
rather indirect and depend on certain assumptions and/or interaction models.
The most bias free mass estimator is probably a measurement of the depth of the
shower maximum \Xmax, preferably with an imaging telescope such as
fluorescence detectors.  The best way to infer the mass is to measure \Xmax
distributions, rather than average values only. However, unfortunately, also
the interpretation of the measured values depends on hadronic interaction
models used in air shower simulations.

\begin{figure}\centering
 \epsfig{width=0.85\textwidth, file=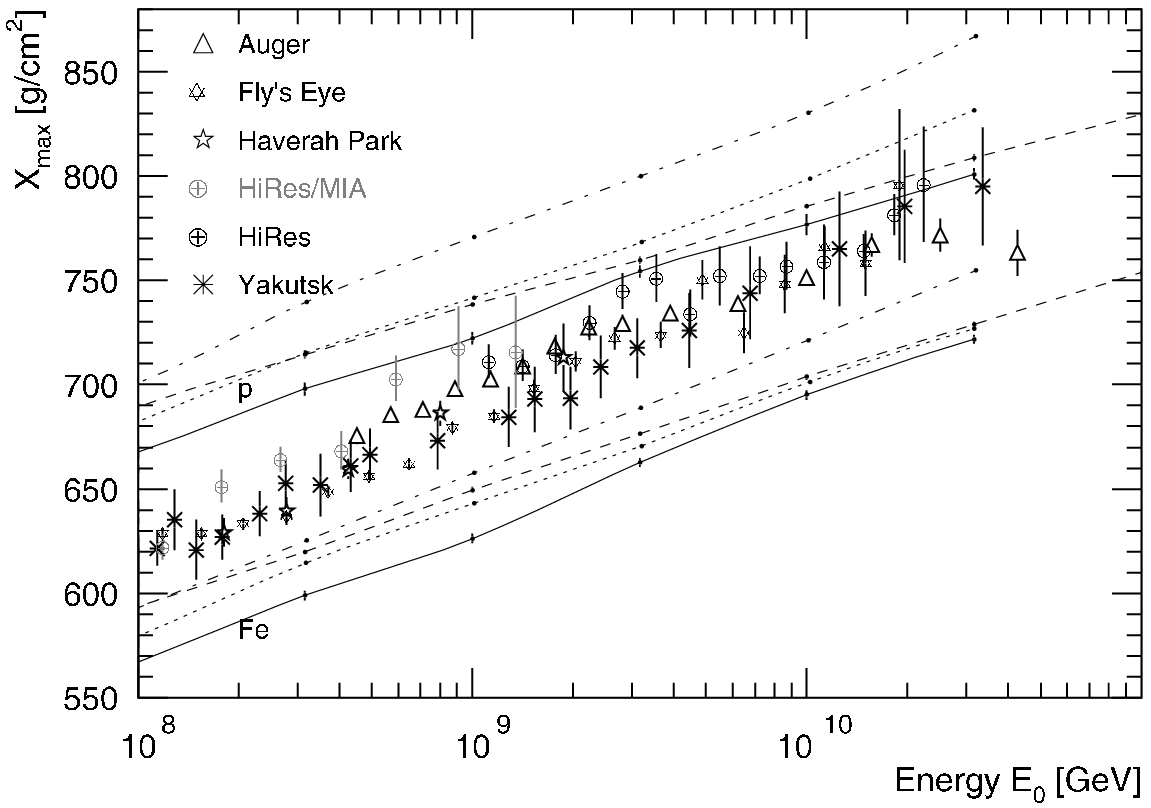}
 \epsfig{width=0.85\textwidth, file=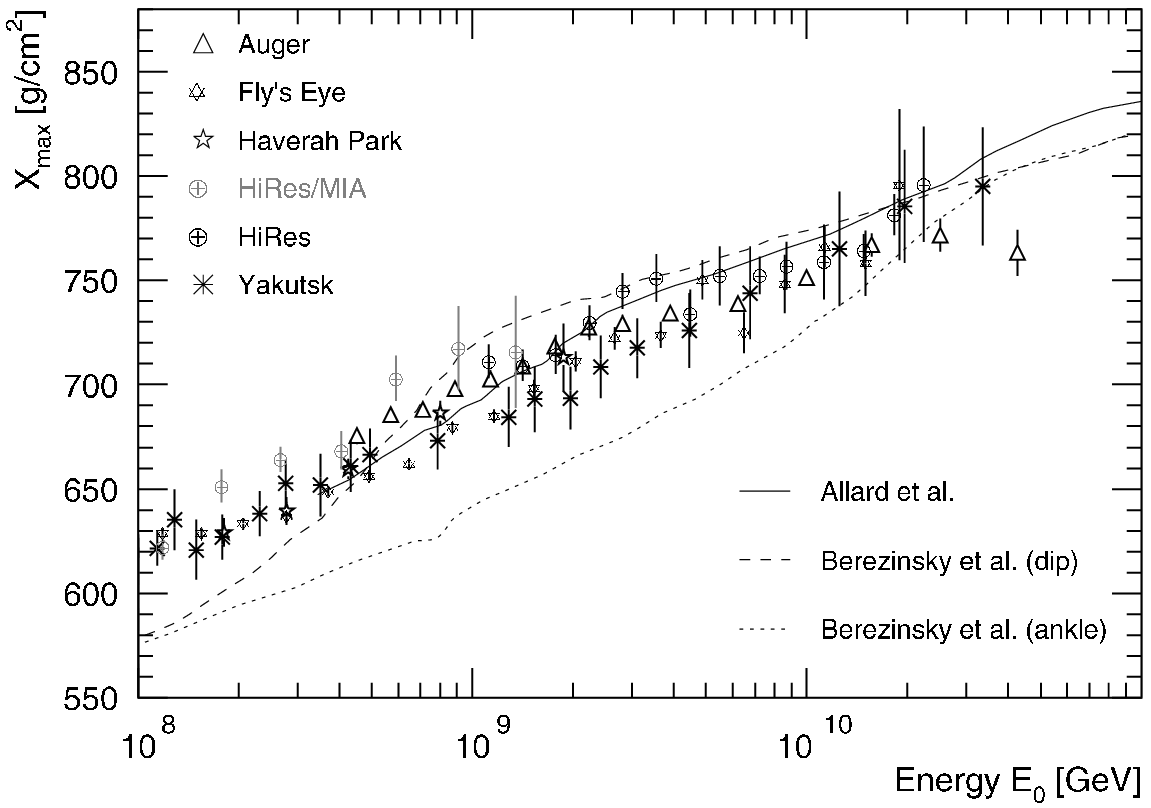}
 \caption{Average depth of the shower maximum \Xmax as function of energy as
          measured by the
	  \PAO \cite{augerxmaxicrc07}, as well as the
          Fly's Eye \cite{flyseye},
          Haverah Park \cite{haverahpark00},
          HiRes/MIA \cite{hires00},
          HiRes \cite{hiresxmax}, and
          Yakutsk \cite{yakutsk} experiments.
	  \TTop: measured values are compared to predictions for primary
	  protons and iron nuclei for different hadronic interaction models
	  QGSJET 01 \cite{qgsjet} (\line),
	  QGSJET II-3 \cite{qgsjet2} (\dashed),
	  SYBILL 2.1 \cite{sibyll21} (\dotted), and
	  DPMJET 2.55 \cite{dpmjet} (\dashdot).
	  \BBottom: comparison to astrophysical models according to
	  \cite{allardxmx} (\line) as well as
	  \cite{berezinskyxmx}, for the latter a  dip (\dashed) and an ankle
	  (\dotted) scenario are distinguished.
	  \label{xmax}}
\end{figure}

The average depth of the shower maximum registered by several experiments is
plotted in \fref{xmax} as function of energy. In the top panel the data are
compared to predictions of air shower simulations for primary protons and iron
nuclei, using different hadronic interaction models, namely QGSJET~01
\cite{qgsjet}, QGSJET~II-3 \cite{qgsjet2}, SYBILL~2.1 \cite{sibyll21}, and
DPMJET~2.55 \cite{dpmjet}. The models yield differences in \Xmax of order of
30~\gcm2 for iron nuclei and $\approx50$~\gcm2 for proton induced showers.  An
overall trend seems to be visible in the data, the measured values seem to
increase faster with energy as compared to the model predictions. This implies
that the composition becomes lighter as function of energy.  Through
interactions with the cosmic microwave background heavy nuclei are expected to
break up during their propagation through the Universe (GZK effect) and a light
composition is expected at the highest energies. However, e.g.\ the Auger data
at the highest energies correspond to a mixed composition for all models
displayed.

In the bottom panel of \fref{xmax} the measured values are compared to
predictions of astrophysical models of the origin of high-energy cosmic rays.\\
The propagation of high-energy cosmic rays in extragalactic turbulent magnetic
fields is considered in \cite{allardxmx}. The average \Xmax values are shown for
a case, assuming a mixed source composition with an injection spectrum $\propto
E^{-2.4}$, a continuous distribution of the sources, and no extragalactic
magnetic field. Other cases studied deliver similar results in \Xmax, for
details see \cite{allardxmx}.\\
Different scenarios for the transition from galactic to extragalactic cosmic
rays are discussed in \cite{berezinskyxmx}.  Two scenarios are distinguished, a
'dip' model in which the galactic and extragalactic fluxes equal at an energy
below $10^{18}$~eV and an 'ankle' approach in which both components have equal
fluxes at an energy exceeding $10^{19}$~eV. It is proposed that the dip is a
consequence of electron-positron pair production, see \sref{propagation}. For
the 'dip' model a source spectrum $\propto E^{-2.7}$ is assumed and a spectrum
$\propto E^{-2}$ for the ankle approach. Cosmic rays have been propagated
through an extragalactic magnetic field of 1~nG. The resulting average \Xmax
values, based on simulations using the interaction code QGSJET~01 are displayed
in the figure.\\
The figure illustrates that we are entering an era where it should be possible
to distinguish between different astrophysical scenarios.

Of great interest is also whether other species than atomic nuclei contribute
to the \uhe particle flux. 

\subsection{Photon Flux Limit}

\begin{figure}\centering
 \epsfig{width=0.8\textwidth, file=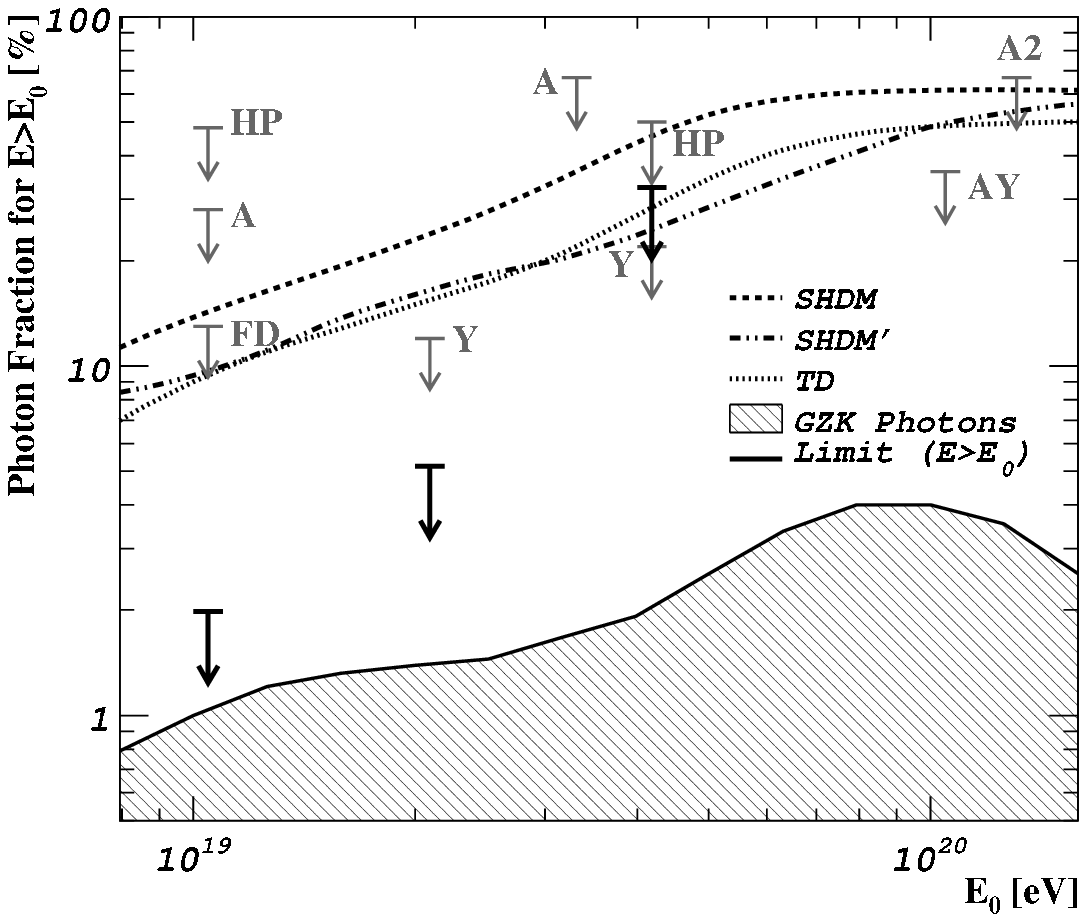}
 \caption{Upper limits on the fraction of photons in the integral cosmic-ray
          flux compared to predictions for GZK photons and top-down scenarios 
	  \cite{augerphoton}. Experimental data are from the Auger surface
	  detectors (arrows) \cite{augerphoton} and 
	  a hybrid analysis (FD) \cite{augerphotonfd},
	  Haverah Park (HP) \cite{haverahphoton},
	  AGASA (A) \cite{agasaphoton,rissephoton},
	  AGASA and Yakutsk (AY) \cite{ayphoton}, as well as
	  Yakutsk (Y) \cite{yakutskphoton}.
	  \label{photon}}
\end{figure}

Air showers induced by primary photons develop an almost pure electromagnetic
cascade. Experimentally they are identified by their relatively low muon
content or their relatively deep shower maximum. Since mostly electromagnetic
processes are involved in the shower development, the predictions are more
reliable and don't suffer from uncertainties in hadronic interaction models.  A
compilation of recent upper limits on the contribution of photons to the
all-particle flux is shown in \fref{photon} \cite{augerphoton}.  The best
photon limits are the latest results of the \PAO \cite{augerphoton} setting
rather strong limits on the photon flux.  They are based on measurements with
the Auger surface detectors, taking into account observables sensitive to the
longitudinal shower development, the signal rise time, and the curvature of the
shower front.  The photon fraction is smaller than 2\%, 5.1\%, and 31\% above
energies of $10^{19}$, $2\cdot10^{19}$, and $4\cdot10^{19}$~eV, respectively
with 95\% confidence level.

In top-down scenarios for high-energy cosmic rays, the particles are decay
products of super-heavy particles. This yields relatively high-fluxes of
photons predicted by such models.  Several predictions are shown in the figure
\cite{aloisiotd,ellistd}. These scenarios are strongly disfavored by
the recent Auger results.

The upper limits are already relatively close to the fluxes expected for
photons originating from the GZK effect \cite{gelmini}, shown in the figure as shaded area.

\subsection{Neutrino Flux Limit}

\begin{figure}\centering
 \epsfig{width=0.8\textwidth, file=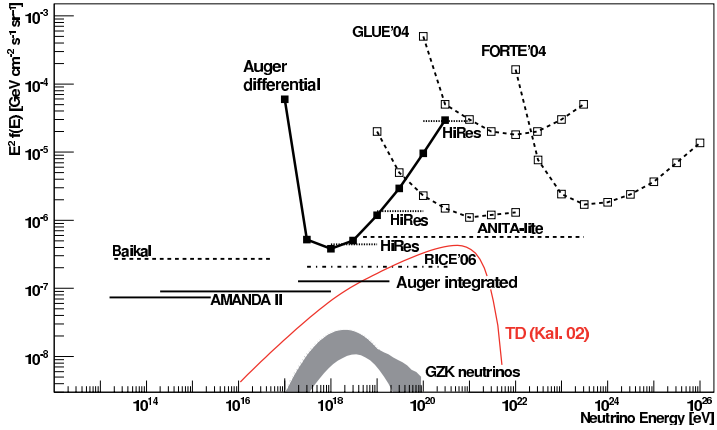}
 \caption{Limits at 90\% confidence level for a diffuse flux of $\nu_\tau$
	  assuming a 1:1:1 ratio of the three neutrino flavors at Earth
	  \cite{augerneutrino,kamperttaup07}. The experimental results are
	  compared to predictions for GZK neutrinos and a top-down model
	  \cite{kalashev}.
	  \label{neutrino}}
\end{figure}

The detection of \uhe cosmic neutrinos is a long standing
experimental challenge. Many experiments are searching for such neutrinos, and
there are several ongoing efforts to construct dedicated experiments to detect
them \cite{halzenrpp,falckenar}. Their discovery would open a new window to the
Universe \cite{becker}.  However, so far no \uhe neutrinos have
been detected. 
\footnote{Neutrinos produced in air showers (atmospheric neutrinos)
   \cite{superkatmospheric}, in the sun \cite{superksolar,sno04}, and during
   super nova 1987A \cite{hirata,bionta} have been detected, but are at
   energies much below our focus. }

As discussed above (\sref{sup}) \UHECR are expected to
be accompanied by \uhe neutrinos.  The neutrinos are produced with
different abundances for the individual flavors, e.g.\ pion decay leads to a
ratio $\nu_e:\nu_\mu=2:1$. However, due to neutrino oscillations the ratio
expected at Earth is $\nu_\tau:\nu_\mu:\nu_e=1:1:1$.

To discriminate against the huge hadronic background in air shower detectors,
neutrino candidates are identified as nearly horizontal showers with a
significant electromagnetic component.
The \PAO is sensitive to Earth-skimming tau-neutrinos that
interact in the Earth's crust. Tau leptons from $\nu_\tau$ charged-current
interactions can emerge and decay in the atmosphere to produce a nearly
horizontal shower with a significant electromagnetic component.  Recent results
from the \PAO together with upper limits from other
experiments are presented in \fref{neutrino}.  Assuming an $E_\nu^{-2}$
differential energy spectrum Auger derives a limit at 90\% confidence level of
$E_\nu^2 \, \mbox{d}N_{\nu_\tau}/\mbox{d}E_\nu < 1.3 \cdot 10^{-7}$~GeV
cm$^{-2}$ s$^{-1}$ sr$^{-1}$ in the energy range between $2\cdot10^{17}$ and
$2\cdot 10^{19}$~eV.

According to top-down models for \UHECR a large flux of
\uhe neutrinos is expected. As an example, the predictions of a
model \cite{kalashev} are shown in the figure as well. This model is
disfavored by the recent upper limits.  It should also be noted that the
current experiments are only about one order of magnitude away from predicted
fluxes of GZK neutrinos (cosmogenic neutrinos).

\section{Arrival Direction}\label{direction}

The arrival directions of cosmic rays provide an important observable to
investigate the sources of these particles. Since charged particles are
deflected in magnetic fields, the cosmic-ray flux observed at Earth is highly
isotropic. A significant evidence for an anisotropy in the arrival directions
would be the most direct hint towards possible cosmic-ray sources.
Unfortunately, only limited experimental information is available about both,
galactic \cite{rand,vallee} and extragalactic
\cite{kronberg,grasso-rubinstein} magnetic fields.
Selecting particles at the highest energies limits the field of view to
distances less than 100~Mpc, see \fref{dipgzk} (\rright).
This implies two advantages: the number of source candidates is limited and the
particles are only slightly deflected since they propagate a restricted
distance only.

\subsection{Galactic Center}

The center of our galaxy is an interesting target for cosmic-ray anisotropy
studies. It harbors a massive black hole associated with the radio source
Sagittarius A$^*$ and a supernova remnant Sagittarius A East. Both are
candidates to be powerful cosmic-ray accelerators.  The importance is
underlined by recent discoveries: the HESS experiment has reported the
observation from TeV $\gamma$ rays near the location of Sagittarius A$^*$
\cite{hessgc} and discovered a region of extended emission from giant molecular
clouds in the central 200~pc of the Milky Way \cite{hessgcnature}.

\begin{figure}
 \epsfig{width=0.51\textwidth, file=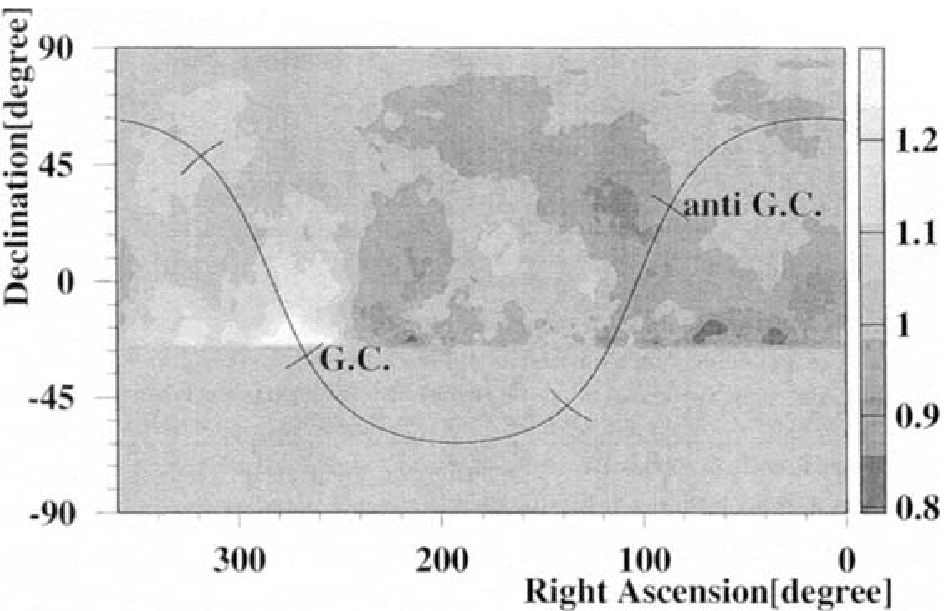}\hspace{\fill}
 \epsfig{width=0.48\textwidth, file=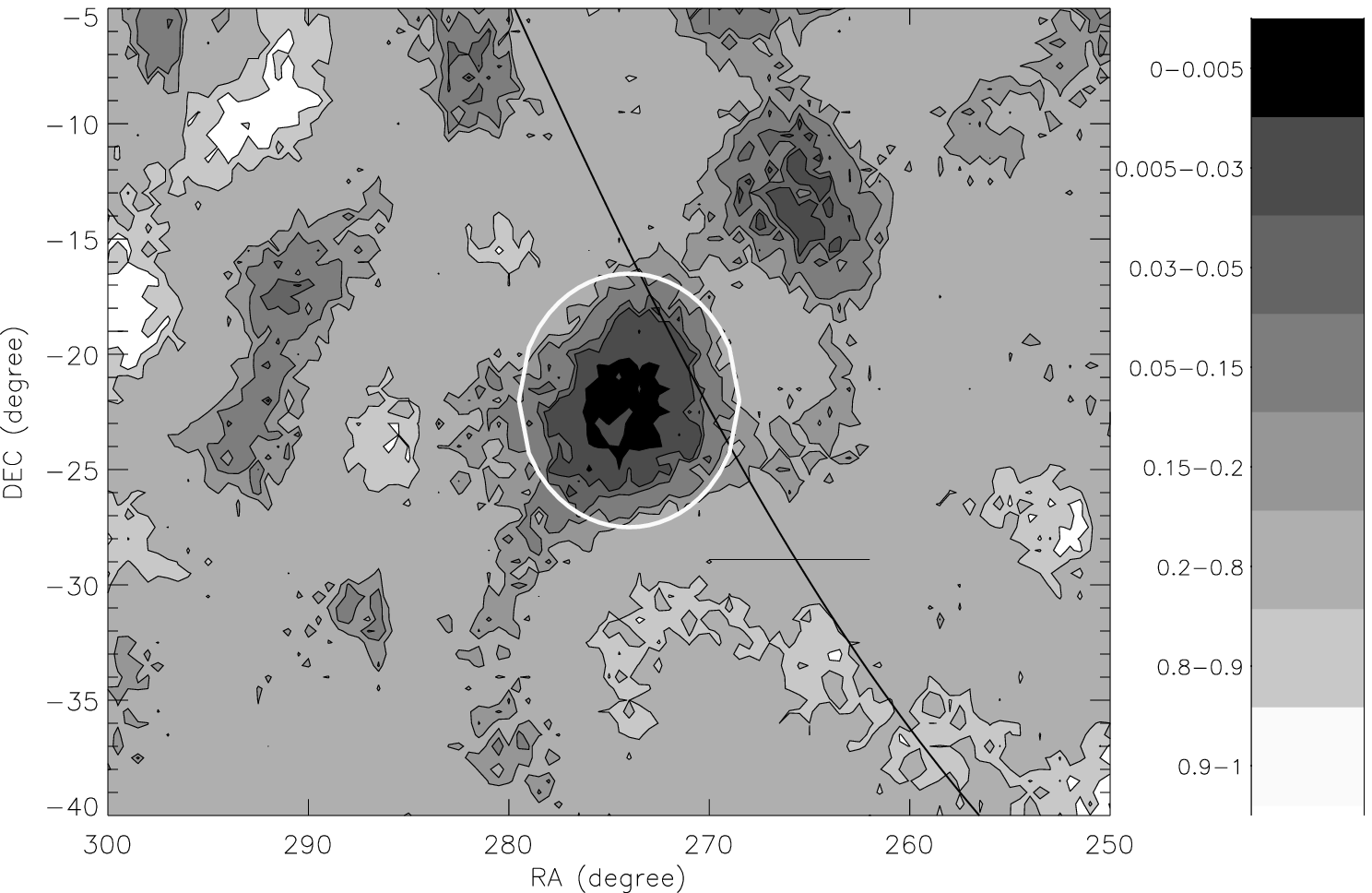}
 \caption{Significance maps of excess/deficit events in equatorial coordinates
	  as measured by the AGASA \cite{agasagc} (\lleft) and SUGAR
	  \cite{sugargc} (\rright) experiments. The lines in both panels
	  indicate the galactic plane. 
	  AGASA: events within a radius of 20\deg are summed up in each bin.
	  SUGAR: The white circle with a radius of 5.5\deg indicates the error
	  for a point source.
          \label{galc}}
\end{figure}

Of particular interest to search for anisotropies in cosmic rays is the region
around $10^{18}$~eV. At these energies the tail of the galactic component might
still contribute significantly to the all-particle spectrum and neutrons from
the galactic center can reach the Earth without decaying. Such neutrons would
not be deflected by magnetic fields
\cite{medinagc,bossagc,aharoniangc,crockergc,grassogc,biermanngc}.

\begin{figure}
 \epsfig{width=0.6\textwidth, file=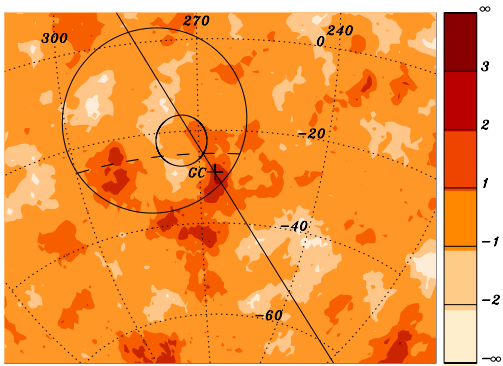}\hspace{\fill}
 \begin{minipage}[b]{0.39\textwidth}
 \caption{Map of cosmic-ray over-density significances near the Galactic Center
	  as measured by the \PAO \cite{augergc}.  The line
	  indicates the galactic plane and "+" marks the Galactic Center. The
	  circles represent the regions of excess events seen by the AGASA and
	  SUGAR experiments.
          \label{augergc}}
 \end{minipage}
\end{figure}

The AGASA experiment has investigated anisotropies in the arrival directions of
cosmic rays at energies around $10^{18}$~eV, see \fref{galc} \cite{agasagc}.
The Galactic Center is just outside the field of view of the experiment.
However, an excess in the Galactic-Center region has been detected.  Also the
SUGAR experiment, located in Australia has reported an excess of events from
the region of the Galactic Center at $10^{18}$~eV, see \fref{galc} (\rright)
\cite{sugargc}. It should be noted that both findings are on the 3 to $4\sigma$
level only.

Recently, data from the \PAO have been searched for anisotropies in the region
of the Galactic Center \cite{augergc}.  A map of resulting cosmic-ray
over-density significances is displayed in \fref{augergc}.  The regions were
AGASA and SUGAR have found an excess are marked in the figure.  With a
statistics much greater than those of previous experiments, it has been
searched for a point-like source in the direction of Sagittarius A.  No
significant excess has been found. Also searches on larger angular scales show
no abnormally over-dense regions. These findings exclude recently proposed
scenarios for a neutron source in the Galactic Center.

\subsection{Clustering of Arrival Directions}

\begin{figure}\centering
 \epsfig{width=0.6\textwidth, file=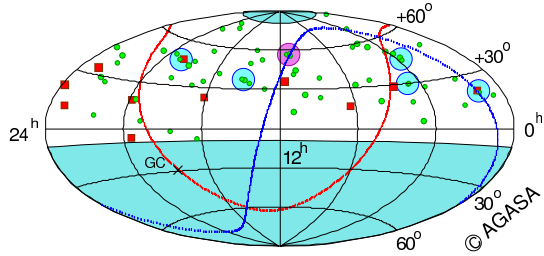}
 \caption{Arrival directions of cosmic rays with energies exceeding
	  $4\cdot10^{19}$~eV in equatorial coordinates as observed by the AGASA
	  experiment.  Red squares  and  green circles  represent cosmic rays
	  with energies exceeding $10^{20}$~eV and  $(4 - 10) \cdot
	  10^{19}$~eV, respectively.  \cite{agasacluster, agasaweb}
	  \label{agasacluster}}
\end{figure}

The AGASA experiment has investigated small-scale anisotropies in the arrival
directions of cosmic rays \cite{agasacluster}. Above an energy of
$4\cdot10^{19}$~eV they have found clusters of events coming from the same
direction, see \fref{agasacluster}. One triplet and three doublets with a
separation angle of 2.5\deg have been reported, the probability to observe
these clusters by a chance coincidence under an isotropic distribution is
smaller than  1\%.

The HiRes experiment has found no significant clustering at any angular scale
up to 5\deg for energies exceeding 10~EeV \cite{hirescluster}.
Combining data from the AGASA, HiRes, SUGAR, and Yakutsk experiments at
energies above 40~EeV a hint for a correlation has been found at angular scales
around 25\deg \cite{kachelriesscluster}.

\begin{figure}
 \epsfig{width=0.6\textwidth, file=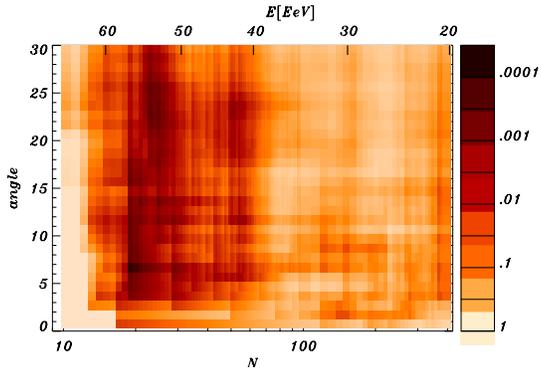}\hspace{\fill}
 \begin{minipage}[b]{0.37\textwidth}
 \caption{Autocorrelation scan for events with energies above $10^{20}$~eV
	  recorded with the \PAO \cite{augerautokorr07}. The chance probability
	  is shown as function of separation angle and threshold energy. 
	  \label{augerautokorr}}
 \end{minipage}
\end{figure}

Also the data of the \PAO have been searched for clustering in the arrival
directions \cite{augerautokorr07}.  The autocorrelation function has been
analyzed adopting a method, in which a scan over the minimum energy $E$ and the
separation angle $\Theta$ is performed \cite{finley}.  For each value of $E$
and $\Theta$ a chance probability is calculated by generating a large number of
isotropic Monte Carlo simulations of the same number of events, and computing
the fraction of simulations having an equal or larger number of pairs than the
data for those parameters. The result is depicted in \fref{augerautokorr},
showing the probability as function of separation angle and threshold energy.
A broad region with an excess of correlation appears at intermediate angular
scales and large energies. The minimum is found at 7\deg for the 19 highest
events ($E>57.5$~EeV), where eight pairs were observed, while one was expected.
The fraction of isotropic simulations with larger number of pairs at that
angular scale and for that number of events is $P_{min}=10^{-4}$. The chance
probability for this value to arise from an isotropic distribution is
$P\approx2\cdot10^{-2}$.

\subsection{Correlation with BL-Lacs}

Interesting candidates as cosmic-ray sources are BL Lacertae objects. They are
a sub class of blazars, active galaxies with beamed emission from a
relativistic jet which is aligned roughly towards our line of sight.  Several
experiments have searched for correlations of the arrival directions of cosmic
rays with the position of BL Lacs on the sky.

A correlation was found between a subset of BL Lac positions and arrival
directions recored by AGASA with energies exceeding 48~EeV and by the Yakutsk
experiment at energies above 24~EeV \cite{tinyakovjetp}.  This correlation and
further ones as reported in \cite{tinyakovapp,gorbunov} between BL Lacs and
\UHECR registered by the AGASA and Yakutsk experiments were not confirmed by
data of the HiRes experiment \cite{hiresbllac}.  On the other hand, an excess
of correlations was found for a subset of BL Lacs and cosmic rays with energies
above 10~EeV \cite{hiresbllac,gorbunovjetp}.

\begin{figure}\centering
 \epsfig{width=0.95\textwidth, file=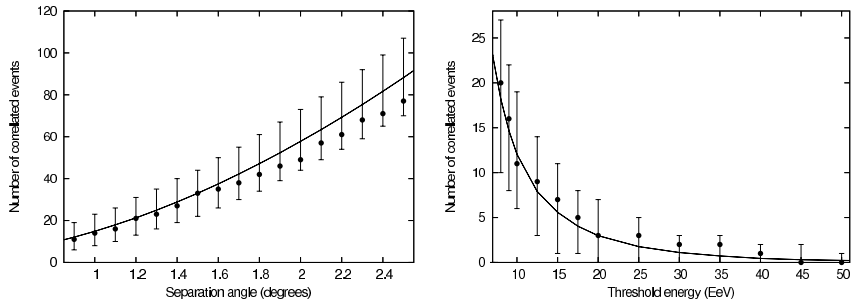}\hspace{\fill}
 \caption{Number of events correlated with confirmed BL Lacs with optical
	  magnitude $m<18$ from the $10^{th}$ edition of the catalog of
	  quasars and nuclei \cite{veroncetty} (points) and average for an
	  isotropic flux (solid line) along with dispersion in 95\% of
	  simulated isotropic sets (bars) \cite{augerbllac07}. As function of
	  the angular separation (for $E>10$~EeV, \lleft) and as function of
	  threshold energy ($\Theta<0.9^\circ$, \rright).
          \label{bllac}}
\end{figure}

In spring 2007 the number of events recorded by the \PAO above 10~EeV was six
times larger than the data used in previous searches. The correlation
hypotheses reported previously have been tested with the Auger data
\cite{augerbllac07}. Since the southern detector of the \PAO sees a different
part of the sky as compared to the AGASA, HiRes, and Yakutsk experiments, only
the 'recipes' could be tested but using different sources on the sky. Non of
the previously reported hypotheses could be confirmed, the chance probabilities
for the different approaches were found to be slightly smaller than 1\%.  The
correlations search has been extended to a broader range of angular scales and
energy thresholds, see \fref{bllac}. It shows the number of correlated events
as function of separation angle (\lleft) and energy threshold (\rright).  The
curves represent expectations for an isotropic flux. The error bars depict the
dispersion within 95\% of simulated isotropic sets. As can be inferred from the
figure, the measured data are compatible with an isotropic distribution and
they do not confirm earlier findings.

\subsection{Correlation with AGN}

\begin{figure}\centering
 \epsfig{width=0.6\textwidth, file=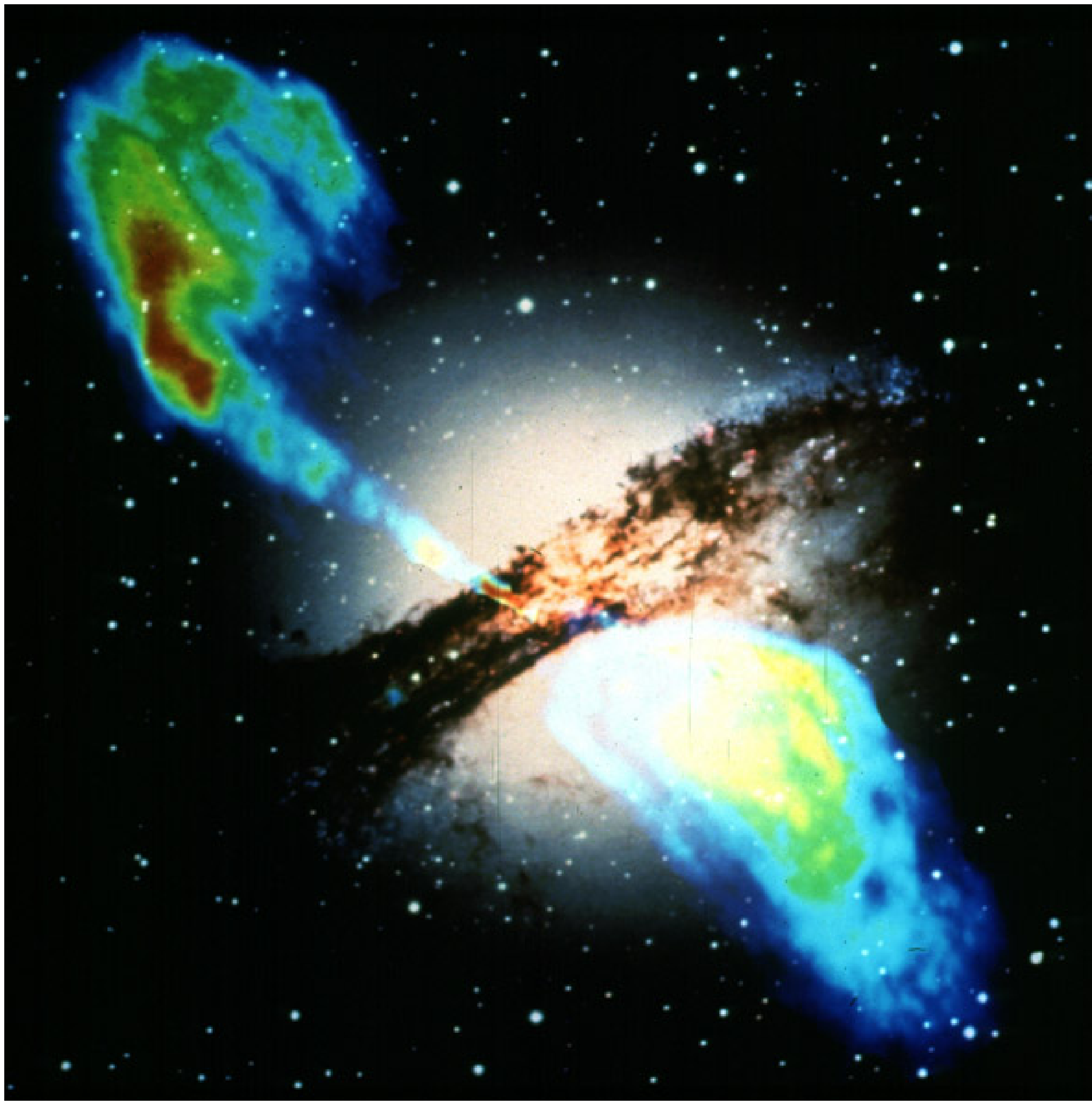}\hspace{\fill}
 \begin{minipage}[b]{0.37\textwidth}
 \caption{Centaurus~A as seen by the Hubble Space Telescope and the VLA radio
          telescope [http://hubblesite.org]. 
	  The radio lobes extend over a scale of about 10\deg along the
	  super-galactic plane. \label{cena}}
 \end{minipage}
\end{figure}

\begin{figure}\centering
 \epsfig{width=0.9\textwidth, file=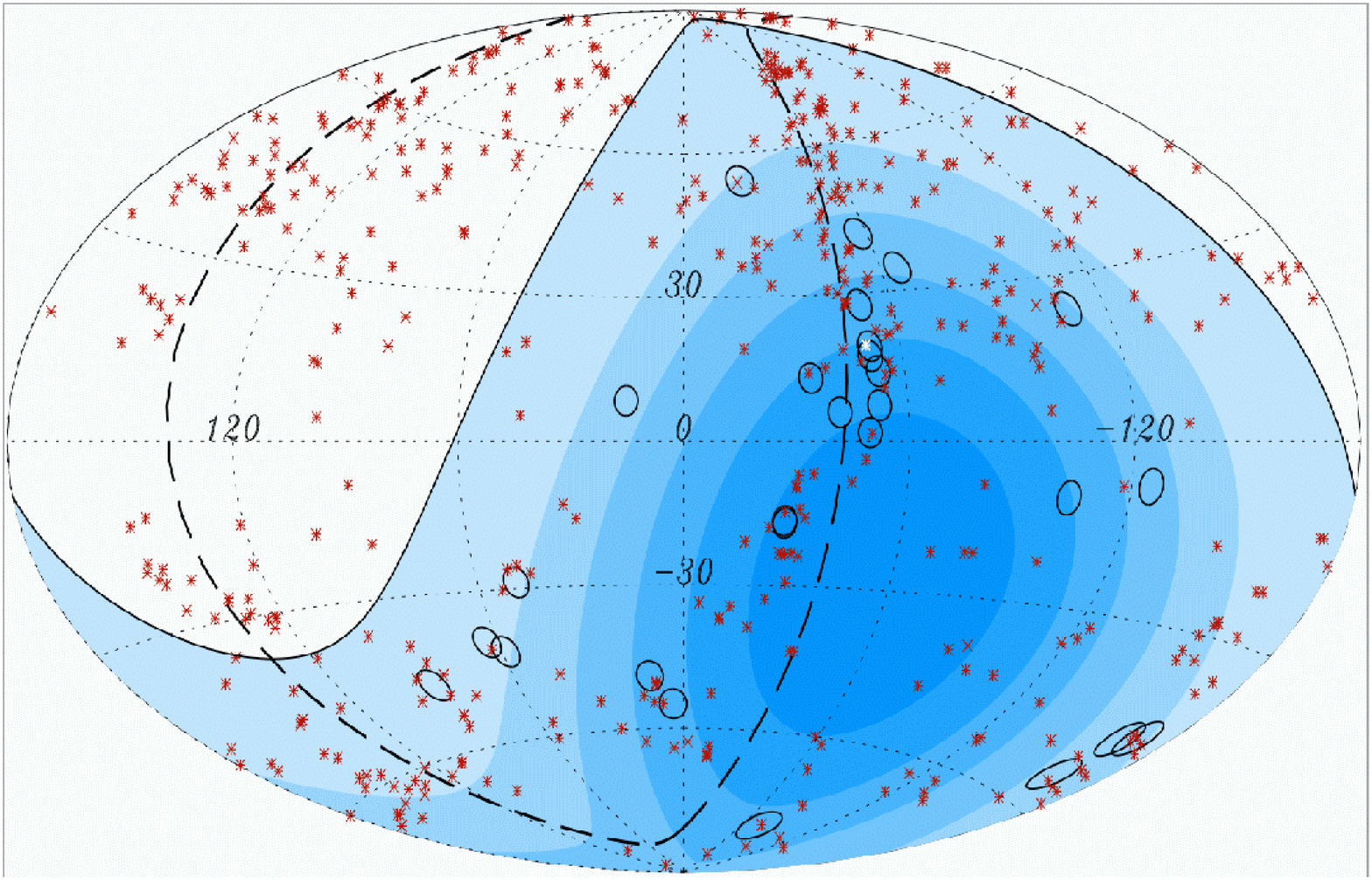}
 \caption{Aitoff projection of the celestial sphere in galactic coordinates
	  with circles of 3.2\deg centered at the arrival directions of 27
	  cosmic rays detected by the \PAO with energies
	  $E>57$~EeV \cite{agnscience}. The positions of AGN with redshift
	  $z<0.018$ ($D<75$~Mpc) from the $12^{th}$ edition of the catalog of
	  quasars and nuclei \cite{veroncetty} are indicated by the asterisks.
	  The solid line draws the border of the field of view of the southern
	  observatory (for zenith angles $\Theta<60^\circ$). Darker color
	  indicates larger relative exposure. The dashed line indicates the
	  super-galactic plane. Centaurus A, one of the closest AGN is marked
	  in white.  
	  \label{agn}}
\end{figure}

Another interesting set of objects to serve as sources of \UHECR are Active
Galactic Nuclei (AGN). The radiation from AGN is believed to be a result of
accretion on to the super-massive black hole (with $10^6$ to $10^8$ solar
masses) at the center of the host galaxy.  AGN are the most luminous persistent
sources of electromagnetic radiation in the Universe. An example of an AGN is
shown in \fref{cena}: Centaurus A is with a distance of 3.4~Mpc one of the
closest AGN.  The radio lobes are thought to be the result of relativistic jets
emerging from the central black hole. Different scenarios related to AGN have
been developed, which are supposed to accelerate particles to highest energies,
e.g.\ \cite{biermannstrittmatter,rachen,romero,ostrowskiagn,lyutikov}.

The arrival directions of cosmic rays as measured by the \PAO have been
correlated with the positions of AGN \cite{agnscience,agnlong}.  Data taken
during the construction of the observatory since January 2004 have been
analyzed, corresponding to slightly more than one year of data of the completed
observatory. The angular resolution of the detector is better than 1\deg at
energies above $10^{19}$~eV \cite{avemerida}.  The positions of AGN according
to the $12^{th}$ edition of the catalog of quasars and nuclei
\cite{veroncetty} within a distance $D$ have been used. A scan has been
performed over the distance $D$, a threshold energy $E_{th}$, and the
correlation angle $\Theta$. The best correlation has been found for events with
energies exceeding $E_{th}=57$~EeV, a maximum distance $D=71$~Mpc,
corresponding to a redshift $z=0.017$, and a correlation angle
$\Theta=3.2^\circ$.  With these parameters 20 out of 27 cosmic rays correlate
with at least one of the 442 selected AGN (292 in the field of view of the
observatory).  Only 5.6 are expected, assuming an isotropic flux.  The 27
cosmic rays measured with the highest energies are shown in \fref{agn} together
with the positions of the AGN. Many of the observed correlated events are
aligned with the super-galactic plane. Two events have arrival directions less
than 3\deg away from Centaurus~A.
These results indicate clearly that the arrival directions
of cosmic rays at highest energies are not isotropic.

A cosmic ray with charge $Ze$ that travels a distance $D$ in a regular magnetic
field $B$ is deflected by an angle \cite{agnlong}
\begin{equation}\label{defleq}
 \delta \approx2.7^\circ \frac{60~\mbox{EeV}}{E/Z}
  \left| \int_0^D\left(\frac{dx}{\mbox{kpc}} \times 
         \frac{B}{3~\mu\mbox{G}}\right)\right| .
\end{equation}
Assuming a coherence scale of order $\approx1$~kpc \cite{stanev97} for the
regular component of the galactic magnetic field, the deflection angle is a few
degrees only for protons with energies larger than 60~EeV. This illustrates
that the observed angular correlations are reasonable, but one has to keep in
mind the limited knowledge about galactic magnetic fields.
The angular scale of the observed correlations also implies that intergalactic
magnetic fields along the line of sight to the sources do not deflect
cosmic-ray trajectories by much more than a few degrees.  The root mean square
deflection of cosmic rays with charge $Ze$, traveling a distance $D$ in a
turbulent magnetic field with coherence length $L_c$ is \cite{agnlong}
\begin{equation}\label{rmseq}
 \delta_{rms} \approx 4^\circ \frac{60~\mbox{EeV}}{E/Z}
 \frac{B_{rms}}{\mbox{nG}} \sqrt{\frac{D}{100~\mbox{Mpc}}}
 \sqrt\frac{L_c}{1~\mbox{Mpc}} .
\end{equation}
As information on intergalactic magnetic fields is very sparse, the
correlations observed can be used to constrain models of turbulent
intergalactic magnetic fields. Within the observed volume they should be such
that in most directions
$B_{rms}\sqrt{L_c}\le10^{-9}~\mbox{G}\sqrt{\mbox{Mpc}}$. In the future the \PAO
will collect more data and more than one event per source should be detected.
It should then be possible to use the data itself to set constraints on magnetic
field models.

It should be noted that the findings by the \PAO imply that the sources of the
highest-energy-cosmic rays are spatially distributed like AGN. The actual
acceleration sites could be the AGN itself or other candidates with the same
spatial distribution as AGN.

\section{Discussion and Outlook}
"How do cosmic accelerators work and what are they accelerating?" is one of
eleven science questions for the new century in physics and astronomy
\cite{nrc}.  In the last few years important progress has been made in
measuring the properties of \UHECR. In particular, the results of the \PAO have
significantly contributed to an improvement in understanding the origin of the
highest-energy particles in the Universe.  The discovery of correlations
between the arrival directions of cosmic rays and the positions of AGN was
among the most important scientific breakthroughs in 2007 for several science
media organizations, see
[www.auger.org/news/top\_news\_2007.html].

The must important findings discussed in this overview may be summarized as
follows. The structures in the energy spectrum at highest energies seem to
become more clear. In particular, there seems to be evidence for a steeper
falling spectrum above $4\cdot10^{19}$~eV (\ffref{augere} and \ref{espek}). The
question arises whether this steepening is due to the GZK effect or due to the
maximum energy achieved during the acceleration processes.  The most convincing
evidence for the existence of the GZK effect is provided by the correlations of
the arrival directions with AGN. They occur sharply above an energy of 57~EeV.
At this energy, the flux measured by the \PAO is about 50\% lower than expected
from a power law extrapolation from lower energies, see \fref{augere}. Thus,
there seems to be a connection between the steepening in the spectrum and the
AGN correlation. 

The correlations occur on an angular scale of about 3.2\deg. This indicates
that the particles are deflected marginally only. In turn, this implies they
should be light particles, with low $Z$, see \eref{defleq} and \eref{rmseq}.
However, there is some tension between this expectation and the measurements of
the average depth of the shower maximum \Xmax (\fref{xmax}). The data at
$4\cdot10^{19}$~eV are compatible with a mixed composition. But, since the
correlations occur relatively sharp above 57~EeV, some dramatic change in
composition above this energy can also not be excluded.

The correlation implies that the sources of \uhe particles are in our
cosmological neighborhood ($D<75$).  The GZK horizon, defined as the distance
from Earth which contains the sources that produce 90\% of the protons that
arrive with energies above a certain threshold is 90~Mpc at 80~EeV and 200~Mpc
at 60~EeV \cite{harari}.  There seems to be a slight mismatch between these
numbers and the Auger findings. Shifting upward the Auger energy scale by about
30\%, as indicated by some simulations of the reconstruction procedures
\cite{engelmerida}, a better agreement between the predicted GZK horizon and
energy threshold with the observed data could be achieved \cite{agnlong}.

The biggest uncertainty in the absolute energy scale of the \PAO is the
knowledge of the fluorescence yield.  At present, intensive efforts are
conducted by various groups to precisely determine the fluorescence yield of
electrons in air \cite{proc5fw}. Attention is paid to the dependence of the
yield on atmospheric parameters, like pressure, temperature, and humidity. In
particular, upcoming results from the AIRFLY experiment
\cite{airflyp,priviteramerida} are expected to significantly reduce the
uncertainties of the energy scale for fluorescence detectors.

The correlation between the arrival directions and the positions of AGN sets
constraints on models for the acceleration of \uhe particles.  The results
imply that the spatial distribution of sources is correlated with the
distribution of AGN. Thus, already some scenarios are strongly disfavored.
Ruled out are models proposing sources in our Galaxy, like neutron stars
\cite{blasins}, pulsars \cite{bednarekpulsars}, and black holes \cite{darbh}.
Models for sources in the galactic halo are also ruled out such as top-down
scenarios with decaying super-heavy particles
\cite{berezinskitd,kuzmintd,birkeltd}.  These models are also severely
constraint by the upper limits on the photon flux (\fref{photon}) and the
neutrino flux (\fref{neutrino}). Within the next years measurements of photons
and neutrinos produced in the GZK effect seem to be in reach. Their detection
would be an important and complementary information about the origin and
propagation of \UHECR.

In summary, the acceleration of ultra high-energy particles in AGN seems to be
very attractive, different scenarios have been proposed, e.g.\
\cite{biermannstrittmatter,rachen,romero,ostrowskiagn,lyutikov}.
However, other sources with a similar spatial distribution are not excluded.

With the energy density estimated in \sref{sources} we obtain a total cosmic
ray power of about $9.7\cdot10^{43}$~erg/s within a sphere ($r=75$~Mpc) seen by
the \PAO at the highest energies. The typical power in the jets of AGN is of
order of $10^{44}$ to $10^{46}$~erg/s \cite{kroeding}. If we assume about 10\%
of this power is converted into cosmic rays, about 1 to 10 sources are needed
to sustain the power of the observed extragalactic cosmic ray flux within a
distance of 75~Mpc from Earth.  If the efficiency is slightly smaller, the
number of sources required is correspondingly slightly larger.  \footnote{Based
on statistical arguments a minimum number of sources $\ge61$ has been estimated
\cite{agnlong}.} If the sources of the highest-energy particles are indeed
related to AGN, the number of correlated events seen by the \PAO seems to be of
the right order of magnitude and one expects to see in future more events from
the same sources.

When the number of correlated events found in the Auger data is compared to
expectations for the AGASA and HiRes experiments, one has to be aware of the
different energy scales, see \tref{etab}.  The energy scales of the AGASA and
HiRes experiment are shifted relative to the \PAO by about 42\% and 20\%,
respectively. If the Auger prescription is applied to the data of these
experiments, the energy threshold has to be adjusted correspondingly.

\paragraph{
In the next years} several experiments focus on the exploration of the energy
region of the transition from galactic to extragalactic cosmic rays ($10^{17}$
to $10^{18}$~eV).  The 0.5~km$^2$ KASCADE-Grande experiment \cite{grande} is
taking data since 2004 \cite{chiavassapune}.  The Ice \Cerenkov detector Ice
Cube \cite{icecube} at the South Pole and its 1~km$^2$ surface air shower
detector Ice Top \cite{icetop} are under construction. In January 2008 40 Ice
Cube strings and 40 surface detectors have been deployed, which implies the
set-up is already 50\% completed.  Further experiments are the Telescope Array
\cite{telescopearray} and its low energy extension TALE, as well as extensions
of the \PAO to lower energies \cite{klagesmerida}.  With this new high-quality
data more detailed information will be available on the energy spectrum and
the composition and it should be possible to distinguish between different
scenarios for the transition from galactic to extragalactic cosmic rays.

A promising complementary detection method for high-energy cosmic rays is the
measurement of radio emission from air showers. This method provides
three-dimensional information about air showers, similar to the fluorescence
technique, but with the advantage of a much higher duty cycle.  In the next
years air showers are expected to be detected with the LOFAR radio observatory
\cite{lofar}. An extensive research and development program is conducted in the
Pierre Auger Collaboration with the goal to build a 20~km$^2$ radio antennae
array \cite{vdbergmerida}.

The southern site of the \PAO covers only a part of the whole sky, see
\fref{agn}. Since the distribution of matter in the Universe is different in
the parts seen from the northern and southern hemispheres it is important to
observe the whole sky.  The Northern Auger Observatory is designed to complete
and extend the investigations begun in the South \cite{anorthmerida}.  To
unambiguously identify the sources of the highest-energy cosmic rays requires
collecting many more events in spite of the steeply falling energy spectrum.
The planned Northern site will be located in Southeast Colorado, USA, having an
instrumented area several times the area of Auger South.

The Northern Observatory needs unrestricted support now, it is the next step in
exploring the high-energy Universe in the upcoming years.  With the completed
\PAO, with its Southern and Northern sites operated simultaneously, an exciting
future in astroparticle physics is ahead of us.  It will establish charged
particle astronomy on the whole sky and will provide high-accuracy data to test
astrophysical models of the origin of \UHECR.  In addition, it will improve our
understanding of fundamental physics.  The data will give insight into topics
like the existence of vacuum \Cerenkov radiation, the smoothness of space, and
tests of Lorentz invariance \cite{risseklinkhammer,galavernisigl}.  Already
now, the existing Auger data set stringent limits on theories.

\section*{Acknowledgment}
I would like to thank the organizers of "Cosmic Matter" for the excellent
meeting, bringing together the astroparticle physics and astronomy communities
and for their invitation to give an overview talk. 
Many thanks to John Harton for critically reading the manuscript.
I'm grateful to my colleagues from the \PAO as well as the LOFAR, LOPES,
KASCADE-Grande, and TRACER experiments for fruitful discussions.

{\small
%\bibliographystyle{elsart-harv}
%\bibliography{cr}

}

\vfill

\end{document}